\documentclass[fleqn,usenatbib,usedcolumns]{mnras}

\usepackage{amsmath}
\usepackage{txfonts}

\usepackage[british]{babel} 
\usepackage[T1]{fontenc}
\usepackage{ae,aecompl}

\usepackage{graphicx}	% Including figure files
\usepackage{xspace}
\usepackage{dcolumn}
\graphicspath{{./JWST_paper_plots/}}
\newcommand{\galform}{{\sc{galform}}\xspace}
\newcommand{\grasil}{{\sc{grasil}}\xspace}

\newcommand{\hMsol}{$h^{-1}$~M$_{\sun}$}

\title[Predictions for deep galaxy surveys with \emph{JWST}]{Predictions for deep galaxy surveys with \emph{JWST} from $\Lambda$CDM}

\author[W. I. Cowley et al.]{
William I. Cowley$^{1\mathrm{,}2}$\thanks{E-mail: cowley@astro.rug.nl (WIC)},  
Carlton M. Baugh$^{1}$,
Shaun Cole$^{1}$,
Carlos S. Frenk$^{1}$,\newauthor
Cedric G. Lacey$^{1}$
\\
$^{1}$Institute for Computational Cosmology, Department of Physics, University of Durham, South Road, Durham, DH1 3LE, UK\\
$^{2}$Kapteyn Astronomical Institute, University of Groningen, PO Box 800, NL-9700 AV Groningen, the Netherlands\\
}

\date{Accepted XXX. Received YYY; in original form ZZZ}

\pubyear{2017}

\begin{document}
\label{firstpage}
\pagerange{\pageref{firstpage}--\pageref{lastpage}}
\maketitle

\begin{abstract}
We present predictions for the outcome of deep galaxy surveys with the \emph{James Webb Space Telescope} (\emph{JWST}) obtained from a physical model of galaxy formation in $\Lambda$CDM.  We use the latest version of the \galform\ model, embedded within a new ($800$~Mpc)$^{3}$ dark matter only simulation with a halo mass resolution of $M_{\rm halo}>2\times10^{9}$~$h^{-1}$~M$_{\sun}$.  For computing full UV-to-mm galaxy spectral energy distributions, including the absorption and emission of radiation by dust, we use the spectrophotometric radiative transfer code \grasil.  The model is calibrated to reproduce a broad range of observational data at $z\lesssim6$, and we show here that it can also predict evolution of the rest-frame far-UV luminosity function for $7\lesssim z\lesssim10$ which is in good agreement with observations.  We make predictions for the evolution of the luminosity function from $z=16$ to $z=0$ in all broadband filters on the Near InfraRed Camera (NIRCam) and Mid InfraRed Instrument (MIRI) on \emph{JWST} and present the resulting galaxy number counts and redshift distributions.  Our fiducial model predicts that $\sim1$ galaxy per field of view will be observable at $z\sim11$ for a $10^4$~s exposure with NIRCam. A variant model, which produces a higher redshift of reionization in better agreement with \emph{Planck} data, predicts number densities of observable galaxies $\sim5\times$ greater at this redshift.  Similar observations with MIRI are predicted not to detect any galaxies at $z\gtrsim6$.  We also make predictions for the effect of different exposure times on the redshift distributions of galaxies observable with \emph{JWST}, and for the angular sizes of galaxies in \emph{JWST} bands.   
\end{abstract}


%%%%%%%%%%%%%%%%%%%%%%%%%%%%%%%%%%%%%%%%%%%%%%%%%%

%%%%%%%%%%%%%%%%% BODY OF PAPER %%%%%%%%%%%%%%%%%%
\section{Introduction}
The \emph{James Webb Space Telescope} (\emph{JWST}) is scheduled for launch in October 2018 and is expected to significantly advance our understanding of the high-redshift ($z\gtrsim7$) Universe \citep[e.g.][]{Gardner06}.  Two of its on-board instruments, the Near InfraRed Camera (NIRCam) and the Mid InfraRed Instrument (MIRI), are dedicated to obtaining broadband photometry over the wavelength range $0.7-25.5$~$\mu$m with unprecedented sensitivity and angular resolution.  This wavelength coverage will enable \emph{JWST} to probe the rest-frame UV/optical/near-IR spectral energy distributions (SEDs) of high-redshift ($z\gtrsim7$) galaxies, opening up a hitherto unexplored regime of galaxy formation and evolution.

An early breakthrough in the study of galaxies in the high-redshift Universe came from the identification of galaxies at $z\sim3$ using the Lyman-break technique \citep[e.g.][]{SteidelHamilton93,Steidel96}.  This study took advantage of the break in galaxy SEDs produced at the Lyman limit ($912$~{\AA}) to identify galaxies at $z\sim3$ by searching for `dropouts' in a set of broadband photometric filters.  The significance of this development in the context of galaxy formation and evolution, in particular, the implications for the cosmic star formation rate density and the formation of massive galaxies in the $\Lambda$CDM cosmological model, was discussed in Baugh et al. (\citeyear{Baugh98}, see also Mo \& Fukugita \citeyear{MoFukugita96} and Mo et al. \citeyear{MoMaoWhite99}).  A further advance came with the installation of the Advanced Camera for Surveys (ACS) on the \emph{Hubble Space Telescope} which, using the $z$-band, pushed the Lyman-break technique selection to $z\sim6$ \citep[e.g.][]{Bouwens03,Stanway03}.  At these redshifts the Lyman-break technique makes use of the fact that neutral hydrogen in the intergalactic medium (IGM) effectively absorbs radiation with wavelengths shorter than the Lyman~$\alpha$ transition ($1216$~{\AA}), resulting in a strong break in the galaxy SED at the observer-frame wavelength of this transition.  Installation of the Wide-Field Camera 3 (WFC3) with near-IR filters increased the number of galaxies that could be identified at $z\sim7$ (e.g. Bouwens et al. \citeyear{Bouwens10}; Wilkins et al. \citeyear{Wilkins10}), pushing the samples of galaxies at these redshifts into the thousands, with a few examples at $z\sim10$.  These advances have been complemented by ground-based telescopes, such as the Visible and Infrared Survey Telescope for Astronomy (VISTA), that typically provide a larger field of view than their space-based counterparts; this has allowed the bright end of the rest-frame far-UV luminosity function to be probed robustly at $z\sim7$ \citep[e.g.][]{Bowler14}.

As observations in the near-IR with \emph{Hubble} have identified the highest-redshift galaxies to date, a wealth of further information regarding galaxy properties at intermediate redshifts ($z\sim3$) has come from surveys with the \emph{Spitzer} Space Telescope in the same wavelength range that will be probed by \emph{JWST} \citep[e.g.][]{Labbe05,Caputi11,Caputi15}, though \emph{JWST} will have greater angular resolution and sensitivity than \emph{Spitzer}.  As a result, \emph{JWST} is expected to greatly increase the number of observed galaxies at $z\gtrsim7$, providing important information about their SEDs which can help characterise their physical properties, whilst also extending observations of the high-redshift Universe towards the first luminous objects at the end of the so-called cosmic dark ages.  

Understandably, in recent years a number of studies have made predictions for galaxy formation at the high redshifts expected to be probed by \emph{JWST}. Numerical hydrodynamical simulations such as the \emph{First Billion Years} simulation suites \citep[e.g.][]{Paardekooper:2013}, the \emph{BlueTides} simulation \citep[e.g.][]{Wilkins:2016,Wilkins:2017}, the \emph{Renaissance} simulations suite \citep[e.g.][]{Xu:2016} and others \citep[e.g.][]{Dayal:2013,Shimizu:2014} have typically focused on the earliest stars and galaxies as potential sources of reionization and have made predictions for the galaxy rest-frame UV luminosity function. These calculations are generally only run to very high redshift ($z\gtrsim6$), as the computational expense of adequately resolving the physical processes involved becomes prohibitive towards later times. As such, there is considerable uncertainty as to whether such simulations would be able to reproduce the galaxy population at $z=0$. It should be noted, however, that some cosmological hydrodynamical simulations are able to reproduce the galaxy population at $z=0$ \citep[e.g.][]{Vogelsberger14,Schaye15}. 

Simple empirical models \citep[e.g.][]{BehrooziSilk:2015, Mason:2015, Mashian:2016} have also been used to make predictions for the high-redshift galaxy rest-frame UV luminosity function. These models are much less computationally expensive than the hydrodynamical schemes mentioned above and as such can be run to $z=0$. However, they ignore most of the physical processes of galaxy formation and instead rely on arbitrary scalings to compute a small number of galaxy properties from those of the host halo. As such they have a limited predictive power and a physical interpretation of their predictions is foregone. Nevertheless, these models can reproduce evolution of the rest-frame far-UV luminosity function in reasonable agreement with observations for $z\lesssim8$ (though they are often calibrated on these data at some redshifts), and suggest small numbers of galaxies will be observable with future \emph{JWST} galaxy surveys at $z\gtrsim10$.

A powerful technique for studying the formation and evolution of galaxies is semi-analytical modelling (see the reviews by Baugh \citeyear{Baugh06} and Benson \citeyear{Benson10}). In such models, the complex physical processes of galaxy formation are fully accounted for and are described by simplified prescriptions that are based on either theoretical arguments or observational or simulation data. This makes the galaxy formation and evolution calculation more computationally tractable, whilst still encapsulating its intrinsic complexity. The free parameters introduced as a result of these simplified prescriptions are then calibrated against a predetermined set of observational data, often requiring that any viable model reproduce the galaxy population observed at $z=0$. Once this has been done the model is fully specified and can be used to make genuine predictions for a wide range of other observable properties at any redshift. An advantage of semi-analytical models is that their predictions can then be readily interpreted in terms of the modelling and interplay of the physical processes involved, and comparing their predictions to observational data provides a test of our understanding of these processes.

Clay et al. (\citeyear{Clay:2015}) made predictions for the evolution of the rest-frame far-UV luminosity function for $z\sim4-7$ using the semi-analytical model of \cite{Henriques15}. However, this model underpredicted the bright end of the observed luminosity function and relied on an \emph{ad hoc} scaling with redshift of the dust optical depth in galactic discs. Liu et al. (\citeyear{Liu:2016}) achieved a better fit to the observed rest-frame far-UV luminosity function using the semi-analytical model {\sc meraxes} \citep{Mutch:2016}. However, this model only provides predictions for $z\gtrsim5$ and does not account for feedback from an active galactic nucleus (AGN) [though see \cite{Qin:2017} for an updated version of this model that addresses these shortcomings]. Additionally, neither of these works attempt to model dust emission and thus their predictions are restricted to the rest-frame UV/optical/near-IR.       

Here we present theoretical predictions for deep galaxy surveys with \emph{JWST} NIRCam and MIRI, in the form of luminosity functions, number counts and redshift distributions from a semi-analytical model of hierarchical galaxy formation within $\Lambda$CDM \citep{Lacey16}. The model provides a physically-motivated computation of galaxy formation and evolution from $z\gtrsim20$ to $z=0$. For computing galaxy SEDs the model is coupled with the spectrophotometric code \grasil \citep{Silva98}, which takes into account the absorption and re-emission of stellar radiation by interstellar dust by solving the equations of radiative transfer in an assumed geometry. This broadens the predictive capability of the model to the full wavelength range that will be probed by \emph{JWST}. The Lacey et al. model is calibrated to reproduce a broad range of observational data at $z\lesssim6$, these include the optical and near-IR luminosity functions at $z=0$, the evolution of the rest-frame near-IR luminosity functions for $z=0-3$, far-IR/sub-mm galaxy number counts and redshift distributions, and the evolution of the rest-frame far-UV luminosity function for $z=3-6$. The predictions of this model presented in this work thus represent an exciting opportunity to test the modelling and interplay of the physical processes of galaxy formation against \emph{JWST} observations at higher redshifts than those at which the model was calibrated. At the same time, they can potentially inform future \emph{JWST} galaxy survey strategies. 

A shortcoming of the fiducial Lacey et al. model, however, is that it does not reproduce the reionization redshift of $z=8.8_{-1.4}^{+1.7}$ inferred from cosmic microwave background (CMB) data by \cite{PlanckCollab15}.  This is an important constraint for high-redshift predictions of the galaxy population.  The model produces too few ionizing photons at early times, reionizing the Universe at $z=6.3$ \citep{HouJun16}.  

A simple and effective solution to this shortcoming was proposed by \cite{HouJun16} who, motivated by the dynamical supernova (SN) feedback model of \cite{Lagos13}, allowed the strength of SN feedback in the Lacey et al. (\citeyear{Lacey16}) model to vary as a function of redshift. Reducing the strength of SN feedback at high redshift meant that the model could produce more ionizing photons at this epoch.  The evolving feedback also enabled this model to reproduce the $z=0$ luminosity function of the Milky Way satellites, as well as their metallicity--stellar mass relation.  These further successes in matching observational data do not come at the expense of the agreement of the model with the data against which it was originally calibrated at $z\lesssim6$, but it does introduce new parameters to describe the effects of SN feedback.  

SN feedback is an extremely important physical process in galaxy evolution \citep[e.g.][]{Larson74,WhiteRees78,WhiteFrenk91,Cole91}.  However, its precise details, for example, exactly how energy input from supernovae (SNe) should couple to the interstellar medium (ISM), are still poorly understood.  This is mainly due to the difficulty of fully resolving individual star-forming regions in hydrodynamical simulations spanning a cosmologically significant time period and volume \citep[e.g.][]{Vogelsberger14,Schaye15}.  It is hoped that comparing the predictions of phenomenological models of SN feedback, such as those presented here, with future observations from \emph{JWST}, will lead to a greater understanding of the efficiency of this crucial process.
   
This paper is structured as follows: in Section~\ref{sec:Model} we present some of the pertinent details of our galaxy formation model and the evolving feedback variant, the radiative transfer code used for the computation of UV-to-mm galaxy SEDs and some information regarding the coupling of these two models. In Section~\ref{sec:Results} we present our main results\footnote{Some of the model data presented here will be made available at \url{http://icc.dur.ac.uk/data/}.  For other requests please contact the first author.}; these include galaxy luminosity functions, number counts and redshift distributions for varying exposures, and angular sizes in each of the NIRCam and MIRI broadband filters.  We also present predictions for the evolution of some of the physical properties of the model galaxies (e.g. stellar masses, star formation rates) and compare some model predictions to available high-redshift ($z\gtrsim7$) observational data. We conclude in Section~\ref{sec:conclusion}. A brief discussion of the dependence of our high-redshift predictions on some of our model assumptions is given in Appendix~\ref{sec:discussion}. 

Throughout we assume a flat $\Lambda$CDM cosmology with cosmological parameters consistent with recent \emph{Planck} satellite results \citep{PlanckCollab15}\footnote{$\Omega_{\rm m}=0.307$, $\Omega_{\Lambda}=0.693$, $h=0.678$, $\Omega_{\rm b}=0.0483$, $\sigma_{8}=0.829$}.  All magnitudes are presented in the absolute bolometric (AB) system \citep{Oke74}.  
\section{The Theoretical Model}
\label{sec:Model}
In this Section we introduce our galaxy formation model, which combines a dark matter only $N$-body simulation, a semi-analytical model of galaxy formation (\galform) and the spectrophotometric radiative transfer code \grasil \citep{Silva98} for computing UV-to-mm galaxy SEDs.
\subsection{GALFORM}
The Durham semi-analytic model of hierarchical galaxy formation, \galform, was introduced in \cite{Cole00}, building on ideas outlined earlier by \cite{WhiteRees78}, \cite{WhiteFrenk91} and \cite{Cole94}.  Galaxy formation is modelled \emph{ab initio}, beginning with a specified cosmology and a linear power spectrum of density fluctuations, and ending with predicted galaxy properties at different redshifts.

Galaxies are assumed to form from baryonic condensation within the potential wells of dark matter halos, with their subsequent evolution being controlled in part by the merging history of the halo. Here, these halo merger trees are extracted directly from a dark matter only $N$-body simulation \citep[e.g.][]{Helly03,Jiang14} as this approach allows us to predict directly the spatial distribution of the galaxies.  We use a new ($800$~Mpc)$^{3}$ Millennium-style simulation \citep{Springel05} with cosmological parameters consistent with recent \emph{Planck} satellite results \citep{PlanckCollab15}, henceforth referred to as P--Millennium (Baugh et al. in preparation; McCullagh et al. \citeyear{McCullagh:2017}). This large volume (800~Mpc)$^3$ gives the bright end of our predicted luminosity functions a greater statistical precision than could be achieved using dark matter only simulations with a better halo mass resolution but smaller volume. 

The halo mass resolution of this simulation is $2.12\times10^{9}$~$h^{-1}$~M$_{\sun}$, where a halo is required to have at least $20$ dark matter particles and is defined according to the `DHalo' algorithm \citep{Jiang14}.  This mass resolution is approximately an order of magnitude better than previous dark matter simulations that were used with this galaxy formation model. For example, the MR7 simulation \citep{Springel05,Guo13} in which the \cite{Lacey16} model was originally implemented had a halo mass resolution of $1.87\times10^{10}$~{\hMsol}.  This improved resolution is particularly important for predictions of the high-redshift Universe where, due to the hierarchical nature of structure formation in $\Lambda$CDM, galaxy formation takes place in lower mass halos.  This halo mass resolution is in the regime where ignoring baryonic effects on the dark matter, an implicit assumption of the semi-analytical technique, is still a reasonable one. The `back-reaction' due to baryonic effects, such as feedback processes, on the dark matter is expected to reduce the mass of dark matter halos by only $\sim30$~per~cent at the mass limit of the P--Millennium simulation \citep[e.g.][]{Sawala13}. 

We have tested that the results presented in this paper have converged with respect to the halo mass resolution used in the P-Millennium simulation and that any artificial features this introduces into our predicted luminosity functions are at luminosities fainter than those studied here. For example, at $z=10$ we find a halo mass resolution `turn-over' in our predicted rest-frame far-UV ($1500$~\AA) luminosity function at $M_{\rm AB}-5\log_{10}h\sim-14$, which is approximately one magnitude fainter than the sensitivity of a $10^{6}$~s exposure with the NIRCam--F150W filter at this redshift.     

Baryonic physics in \galform\ are included as a set of coupled differential equations which track the exchange of mass and metals between between the stellar, cold disc gas and hot halo gas components in a given halo.  These equations comprise simplified prescriptions for the physical processes (e.g. gas cooling, star formation and feedback) understood to be important for galaxy formation. We discuss some of the main features of the model below and refer the interested reader to \cite{Lacey16} for more details.
\subsubsection{The star formation law and stellar initial mass function}   
Star formation in the galactic disc is based on the surface density of molecular gas. Cold disc gas is partitioned into molecular and atomic components based on an empirical relation involving the mid-plane gas pressure, $P$, proposed by \cite{BlitzRosolowsky06} based on observations of nearby galaxies:  
\begin{equation}
R_{\rm mol} = \frac{\Sigma_{\rm mol}}{\Sigma_{\rm atom}} = \left[\frac{P}{P_{0}}\right]^{\alpha_{\rm P}}\rm,
\end{equation}
where $R_{\rm mol}$ is the ratio of molecular to atomic gas; $\alpha_{\rm P}=0.8$ and $P_{0}=1.7\times10^{4}$~cm$^{-3}$~K based on the local observations of \cite{Leroy:2008}. It is assumed that gas and stars are distributed in a exponential disc, the radial scalelength of which is predicted by \galform\ (see Section~\ref{sec:galform_sizes}). The star formation rate surface density is then given by
\begin{equation}
\Sigma_{\rm SFR} = \nu_{\rm SF}\,\Sigma_{\rm mol} = \nu_{\rm SF}\,f_{\rm mol}\,\Sigma_{\rm cold}\rm,
\label{eq:star_formation_law}
\end{equation} 
where $f_{\rm mol}=R_{\rm mol}/(1 + R_{\rm mol})$ and the parameter $\nu_{\rm SF}=0.74$~Gyr$^{-1}$, based on the observations of \cite{Bigiel11}. This expression is then integrated over the whole disc to yield the global star formation rate, $\psi$. For further details of this star formation law we refer the reader to \cite{Lagos11}.  For star formation in the galactic disc a \cite{Kennicutt83} stellar initial mass function (IMF) is assumed.  This IMF is described by $x=0.4$ in $\mathrm{d}N/\mathrm{d}\ln m\propto m^{-x}$ for $m<1$~M$_{\odot}$ and $x=1.5$ for $m>1$~M$_{\odot}$ [for reference, a Salpeter (\citeyear{Salpeter55}) IMF has an unbroken slope of $x=1.35$].

Star formation in bursts, triggered by a dynamical process (see Section~\ref{sec:dyn_process}), takes place in a forming galactic bulge. It is assumed that $f_{\rm mol}\approx1$ and the star formation rate depends on the dynamical timescale of the bulge
\begin{equation}
\psi_{\rm burst} = \nu_{\rm SF,burst}M_{\rm cold,burst}\rm,
\end{equation} 
where $\nu_{\rm SF,burst}=1/\tau_{\star\rm,burst}$ and
\begin{equation}
\tau_{\star\rm,burst} = \max[f_{\rm dyn}\tau_{\rm dyn,bulge},\tau_{\rm burst,min}]\rm.
\end{equation} 
Here $\tau_{\rm dyn,bulge}$ is the dynamical time of the bulge and $f_{\rm dyn}$ and $\tau_{\rm burst,min}$ are model parameters.  This means that for large dynamical times the star formation rate scales with the dynamical time, but has a floor value when the dynamical time of the bulge is short. Here $f_{\rm dyn}=20$ and $\tau_{\rm burst,min}=100$~Myr \citep{Lacey16}.  

For star formation in bursts, it is assumed that stars form with a top-heavy stellar initial mass function (IMF), described by a slope of $x=1$ in $\mathrm{d}N/\mathrm{d}\ln m\propto m^{-x}$. This assumption is primarily motivated by the requirement that the model reproduce the observed far-IR/sub-mm galaxy number counts and redshift distributions \citep[e.g.][see also Fontanot \citeyear{Fontanot:2014} for a study of the effects of IMF variation in semi-analytical models]{Baugh05,Cowley15,Lacey16}. It should be noted that the slope in this new model is much less top-heavy than the one suggested by \cite{Baugh05}, where $x=0$ was assumed. 

The assumption of a top-heavy IMF for starburst galaxies is often seen as controversial. For example, in their review of observational studies \cite{Bastian:2010} argue against significant IMF variation in the local Universe. However, \cite{Gunawardhana:2011} infer an IMF for nearby star-forming galaxies that becomes more top-heavy with increasing star formation rate, reaching a slope of $x\approx0.9$, and a similar IMF slope was inferred for a star-forming galaxy at $z\sim2.5$ by \cite{Finkelstein:2011}. Both of these studies utilise modelling of a combination of nebular emission and broadband photometry to infer an IMF slope. More recently, \cite{Romano:2017} inferred an IMF slope of $x=0.95$ in nearby starburst galaxies through modelling the observed CNO isotopic ratios. Thus whilst the issue of a varying IMF is far from resolved, there are a number of observational studies that support both this assumption and the adopted value of $x=1$.
  
\subsubsection{Feedback processes}
The model includes three modes of feedback from stars and AGN on the galaxy formation process.

\emph{Photoionization feedback}: the IGM is reionized and photo-heated by ionizing photons produced by stars.  This inhibits star formation through (i) preventing gas accretion onto low-mass haloes through an increased IGM pressure and (ii) continued photo-heating reducing the cooling rate of gas already within haloes. Here a simple scheme is implemented that assumes that after the IGM is reionized at a fixed redshift, $z_{\rm reion}$, no cooling of gas occurs in haloes with circular velocities $V_{\rm vir}<V_{\rm crit}$.  Here we assume $z_{\rm reion}=10$ \citep{Dunkley:2009}\footnote{This value of $z_{\rm reion}$ is slightly different to the one predicted by the models, however varying this parameter within the range suggested by the model predictions has a negligible effect on our results [see Fig.~\ref{fig:param_explore}(c)].} and $V_{\rm crit}=30$~km~s$^{-1}$, based on hydrodynamical simulations \citep[e.g.][]{Hoeft:2006,Okamoto:2008}.  Whilst this model is very simple it is based on a self-consistent calculation of reionization in \galform\ described by \cite{Benson:2002}, and it was shown by \cite{Font:2011} to reproduce results from more detailed treatments \citep[e.g.][]{Munoz:2009,Busha:2010} of this process. 

\emph{Supernova (SN) feedback}: the injection of energy into the ISM from SNe ejects gas from the disc to beyond the virial radius of the halo at a rate, $\dot{M}_{\rm eject}$. As SNe are short-lived this rate is  proportional to the star formation rate, $\psi$, according to a `mass loading' factor, $\beta$, such that
\begin{equation}
\dot{M}_{\rm eject} = \beta(V_{\rm c})\,\psi = \left(V_{\rm c}/V_{\rm SN}\right)^{-\gamma_{\rm SN}}\psi\rm.
\label{eq:sn_feedback}
\end{equation}
Here $V_{\rm c}$ is the circular velocity of the disc; $\psi$ is the star formation rate; and $V_{\rm SN}$ and $\gamma_{\rm SN}$ are adjustable parameters.  We assume $V_{\rm SN}=320$~km~s$^{-1}$ \citep{Lacey16} and $\gamma_{\rm SN}=3.4$ (Baugh et al. in preparation, see Section~\ref{sec:changes}). The ejected gas accumulates in a reservoir of mass $M_{\rm res}$, and then falls back within the virial radius at a rate
\begin{equation}
\dot{M}_{\rm return} = \alpha_{\rm ret}\frac{M_{\rm res}}{\tau_{\rm dyn,halo}}\rm,
\label{eq:gas_reincorp}
\end{equation}
where $\tau_{\rm dyn,halo}$ is the halo dynamical time and $\alpha_{\rm ret}=1.0$ (Baugh et al., in preparation, see also Section~\ref{sec:changes}).

\emph{AGN feedback}: the model implements a hot-halo mode AGN feedback, first implemented into \galform\ by \cite{Bower06}. Energy released by the direct accretion of hot gas from the halo onto the supermassive black hole (SMBH) powers relativistic jets that deposit thermal energy into the hot halo gas and thus inhibit further cooling. In the model, gas cooling is turned off if: (i) the gas is cooling quasi-statically (i.e. the cooling time is long compared to the free-fall time), and (ii) the SMBH is massive enough such that the power required to balance the radiative cooling luminosity of the gas is below some fraction of its Eddington luminosity.
      
\subsubsection{Dynamical processes}
\label{sec:dyn_process}
Morphological transitions occur, and starbursts are triggered, through dynamical processes.  These are either galaxy mergers, where the orbit of a satellite galaxy in a dark matter halo has decayed through dynamical friction such that it merges with the central galaxy, or disc instabilities, in which the galactic disc becomes sufficiently self-gravitating that it is unstable to bar formation [using the criterion of Efstathiou et al. (\citeyear{Efstathiou:1982}), which is based on simulations of isolated disc galaxies].   

Major galaxy mergers (and minor mergers above a baryonic mass ratio) and all disc instabilities trigger bursts of star formation.  In these, all of the cold gas in the disc is transferred to a forming bulge/spheroid and forms stars according to the star formation law for bursts and assuming a top-heavy IMF as is described earlier. 

When we refer to starburst galaxies throughout this paper, we are referring to this dynamically triggered star formation rather than, for example, a galaxy's position on the specific star formation rate--stellar mass plane.  This distinction is discussed in more detail in \cite{Cowley16SEDs}.     
\subsubsection{Galaxy sizes}
\label{sec:galform_sizes}
In \galform\ it is assumed that a disc with an exponential radial profile is formed from cold gas once it has had sufficient time to cool and fall to the centre of the dark matter halo potential well. The size of the disc is calculated by assuming conservation of angular momentum and centrifugal equilibrium \citep{Cole00}. 

Galaxy bulges/spheroids are assumed to have a projected $r^{1/4}$ density profile and are formed through a dynamical process, either a disc instability or a galaxy merger. The size of the bulge is determined by the conservation of energy for the components involved i.e. baryons and dark matter in the disc and bulge of the galaxies \citep{Cole00}.
\subsubsection{Changes to the Lacey et al. (\citeyear{Lacey16}) model}
\label{sec:changes}
\begin{table*}
\centering
\caption{Changes between parameter values presented in Lacey et al. (\citeyear{Lacey16}) and those used in this work (and discussed further in Baugh et al. in preparation).  The galaxy formation parameters are listed in the bottom part of the table.}
\begin{tabular}{llll}\hline
Parameter & Description & Lacey et al. (\citeyear{Lacey16}) & This work\\ \hline
\multicolumn{2}{l}{Cosmological parameters} & Komatsu et al. (\citeyear{Komatsu11}) & \emph{Planck} Collaboration (\citeyear{PlanckCollab15})\\
$\Omega_{\rm m}$ & Matter density &$0.272$&$0.307$ \\
$\Omega_{\Lambda}$ & Vacuum energy density &$0.728$&$0.693$ \\
$\Omega_{\rm b}$ & Baryon density &$0.0455$&$0.0483$ \\
$h$   & Hubble Parameter &$0.704$&$0.678$\\
$\sigma_{8}$ & Fluctuation amplitude &$0.810$&$0.829$\\ \hline
\multicolumn{4}{l}{$N$-body simulation parameters}\\  
$M_{\rm halo,min}$ & Minimum halo mass & $1.87\times10^{10}$~$h^{-1}$~M$_{\sun}$ & $2.12\times10^{9}$~$h^{-1}$~M$_{\sun}$\\ \hline
\multicolumn{2}{l}{Galaxy merger timescale}&Jiang et al. (\citeyear{CYJiang08})&Simha \& Cole (\citeyear{SimhaCole16})\\ \hline
\multicolumn{4}{l}{Galaxy formation parameters} \\
$\alpha_{\rm ret}$&Gas reincorporation timescale factor&$0.64$&$1.00$\\
$\gamma_{\rm SN}$&Slope of SN feedback mass loading&$3.2$&$3.4$\\ \hline  
\end{tabular}
\label{table:param_changes}
\end{table*}     
This work assumes different cosmological parameters from those assumed by \cite{Lacey16}, and utilises an  $N$-body simulation with a better halo mass resolution. The model used here also incorporates an improved prescription for the merger timescale of satellite galaxies \citep{SimhaCole16}, which was first introduced into \galform\ by \cite{Campbell:2015}, but was not considered by Lacey et al. This new treatment accounts for the effects of both dynamical friction and tidal stripping on the sub-halo and thus more closely follows the underlying $N$-body simulation than the analytical prescription used in \galform\ previously \citep{LaceyCole:1993,CYJiang08}. Additionally, the earlier prescription for the merger timescale resulted in a radial distribution of satellite galaxies that was too centrally concentrated \citep{Contreras:2013}. 

As a result of these changes, it is necessary to adjust some of the galaxy formation parameters in the fiducial model so that it can still reproduce certain pre-specified observational datasets to the desired accuracy. The adjustments will be discussed in more detail in Baugh et al. (in preparation); however, we briefly summarise the main ideas here. A minor reduction in the number of bright galaxies at $z=0$ required the gas reincorporation timescale factor, $\alpha_{\rm ret}$ (equation~\ref{eq:gas_reincorp}), to be increased from $0.64$ to $1.00$, thus returning gas ejected by SN feedback to the hot halo faster. Additionally, the change in the halo mass resolution resulted in the number of faint galaxies being slightly overpredicted, so it was necessary to increase the strength of the SN feedback through increasing the value of the parameter $\gamma_{\rm SN}$ (equation~\ref{eq:sn_feedback}) from $3.2$ to $3.4$ to mitigate this.    

We summarise these minor adjustments to the model presented in \cite{Lacey16} in Table~\ref{table:param_changes}.
\subsection{Evolving supernova feedback and the redshift of reionization}
\label{subsec:EvolFB}
\begin{figure}
\centering
\includegraphics[trim = 0 0 4.135in 0,clip = True,width = \linewidth]{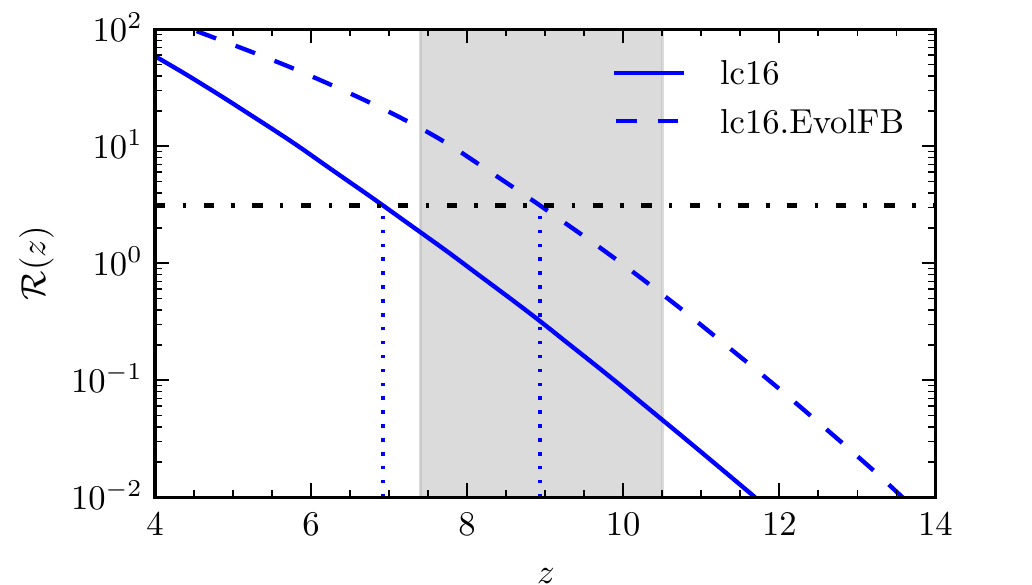}
\caption{Predicted ratio, $\mathcal{R}(z)$, of the total number of ionizing photons produced before redshift $z$ to the total number of hydrogen nuclei, for the fiducial model (solid blue line) and the evolving feedback variant (dashed blue line).  The horizontal black dot-dashed line indicates the ratio at which the IGM is half ionized, $\mathcal{R}_{\rm re,\,half}$.  The grey shaded region indicates the observational estimate of the redshift at which this happens, $z_{\rm re,\,half}=8.8^{+1.7}_{-1.4}$, the $68$~per~cent confidence limit from the \emph{Planck} Collaboration (\citeyear{PlanckCollab15}).  Dotted vertical lines indicate the values of $z_{\rm re,\,half}$ predicted by the models.}
\label{fig:z_reion}
\end{figure}
As mentioned earlier, a shortcoming of the fiducial \cite{Lacey16} model is that it does not reionize the Universe at a redshift as high as implied by recent \emph{Planck} data, as it does not produce enough ionizing photons at early enough times.  Here we discuss the variant feedback model of Hou et al. (\citeyear{HouJun16}) which provides a simple and effective solution to this shortcoming.

In the fiducial \galform\ model, gas outflows due to SN feedback are implemented according to equation~\ref{eq:sn_feedback}. A dynamical model of SN feedback, which followed the evolution of pressurized SNe bubbles in a multiphase ISM was implemented into the \galform\ framework by \cite{Lagos13}. Whilst this SN feedback model is not complete as it only considers gas escaping from the galactic disc, and not from the halo, it suggested that the dependence of the mass loading, $\beta$, solely on galaxy circular velocity may be an oversimplification of this physical process. This standard parametrization of $\beta$ is motivated by the fact that gas outflows should depend on the depth of the gravitational potential well, for which $V_{\rm c}$ is a commonly used proxy. However, it is reasonable to propose that it may also depend on properties such as the gas density, the gas metallicity and the molecular gas fraction. For example, the local gas density and metallicity determine the local gas cooling rate in the ISM and in turn the fraction of the injected SN energy that can be used to launch outflows; and dense molecular gas may not be affected by SNe explosions and thus not ejected in such outflows. These additional physical parameters will evolve with redshift, and may not be well described by a power law with $V_{\rm c}$.

In order to produce more ionizing photons, and thus reionize the Universe earlier than the fiducial model, Hou et al., motivated by the dynamical SN feedback model of \cite{Lagos13}, introduced a break into the power-law parametrisation of the mass loading factor and also a redshift dependence into its normalisation, such that
\begin{equation}
\beta(V_{\rm c},z) =  
\begin{cases}
 [V_{\rm c}/V_{\rm SN}^{\prime}(z)]^{-\gamma_{\rm SN}^{\prime}} & V_{\rm c}\leq V_{\rm thresh} \\
 [V_{\rm c}/V_{\rm SN}(z)]^{-\gamma_{\rm SN}} & V_{\rm c}>V_{\rm thresh}\rm, \\
\end{cases}
\end{equation}  
where $V_{\rm thresh}$ and $\gamma_{\rm SN}^{\prime}$ are additional adjustable parameters [$V_{\rm SN}^{\prime}(z)$ is set by the condition that $\beta$ be a continuous function at $V_{\rm c}=V_{\rm thresh}$].  The redshift evolution of the normalisation is parametrised as
\begin{equation}
V_{\rm SN}(z) =
\begin{cases}
V_{\rm SN2}             & z>z_{\rm SN2}\\
c_{0}\,z+c_{1} & z_{\rm SN2}\leq z\leq z_{\rm SN1}\\
V_{\rm SN1}             & z<z_{\rm SN1}\rm,\\
\end{cases}
\end{equation}
where $V_{\rm SN2}$, $z_{\rm SN2}$ and $z_{\rm SN1}$ are additional adjustable parameters [the constants $c_{0}$ and $c_{1}$ are set by the condition that $V_{\rm SN}(z)$ be a continuous function]. This form parametrizes our ignorance of the precise physical mechanisms at play, whilst allowing for the dependencies of the mass outflow rates on physical properties other than $V_{\rm c}$, as discussed above, to be described. Though we acknowledge that a detailed physical interpretation of this variant feedback model is somewhat lacking, it provides a tractable approximation that is calibrated not only on the reionization redshift suggested by \emph{Planck} data, but also on the luminosity function and metallicity--stellar mass relation of $z=0$ Milky Way satellites. These independent observational data provide strong constraints on the form that the mass-loading factor for SN feedback can take, as is discussed in \cite{HouJun16}.

Here we use the same values for the additional adjustable parameters in this variant feedback model as Hou et al.: $V_{\rm thresh}=50$~km~s$^{-1}$, $\gamma_{\rm SN}^{\prime}=1.0$, $V_{\rm SN2}=180$~km~s$^{-1}$, $z_{\rm SN1}=4$ and $z_{\rm SN2}=8$, without any further calibration, although we remind the reader that the value for $\gamma_{\rm SN}$ is different to the one used by Hou et al. Additionally, we adopt $V_{\rm SN1}=V_{\rm SN}$, as was done by Hou et al.

We show the predicted redshift of reionization for both the fiducial model (lc16) and the evolving feedback variant (lc16.EvolFB) in Fig.~\ref{fig:z_reion}.  Following Hou et al. we calculate the ratio, $\mathcal{R}(z)$, of ionizing photons produced before redshift $z$, to the number density of hydrogen nuclei as
\begin{equation}
\mathcal{R}(z) = \frac{\int_{z}^{\infty}\epsilon(z^{\prime})\,\mathrm{d}z^{\prime}}{n_{\rm H}}\rm,
\end{equation}
where $\epsilon(z^{\prime})$ is the number of hydrogen-ionizing photons produced per unit comoving volume per unit redshift at redshift $z^{\prime}$, and $n_{\rm H}$ is the comoving number density of hydrogen nuclei.  The Universe is assumed to be fully ionized at redshift $z_{\rm re,\,full}$, for which,
\begin{equation}
\mathcal{R}(z_{\rm re,\,full})=\frac{1+N_{\rm rec}}{f_{\rm esc}}\rm,
\end{equation}
where $N_{\rm rec}$ is the mean number of recombinations per hydrogen atom up to reionization, and $f_{\rm esc}$ is the fraction of ionizing photons that can escape into the IGM from the galaxy producing them.  Here we adopt $N_{\rm rec}=0.25$ and $f_{\rm esc}=0.2$ as was done by Hou et al.  This gives a threshold for reionization of $\mathcal{R}(z_{\rm re,full})=6.25$.  

Observations of the CMB \citep[e.g.][]{PlanckCollab15} directly constrain the electron scattering optical depth to recombination, which is then converted to a reionization redshift by assuming a simple model for the redshift dependence of reionization (e.g. Appendix B of Lewis et al., \citeyear{Lewis08}).  The redshift of reionization is commonly expressed in terms of the redshift, $z_{\rm re,\,half}$, at which half of the IGM is reionized.  Here we assume $\mathcal{R}_{\rm re,\,half}=0.5\,\mathcal{R}_{\rm re,\,full}$ as was done by Hou et al.  The value of $\mathcal{R}_{\rm re,\,half}$ is shown as the horizontal dot-dashed line in Fig.~\ref{fig:z_reion}.  We can see that the evolving feedback model predicts $z_{\rm re,\,half}=8.9$, in good agreement with the $68$~per~cent confidence interval inferred from \emph{Planck} satellite data \citep{PlanckCollab15}, $z_{\rm re,\, half}=8.8_{-1.4}^{+1.7}$.  For the fiducial model the reionization redshift turns out to be lower, $z_{\rm re,\,half}=6.9$, which is discrepant by $\sim1.5\sigma$ with the \emph{Planck} data.
\subsection{The Dust Model}
\label{subsec:grasil}
\begin{figure*}
\includegraphics[width=\linewidth]{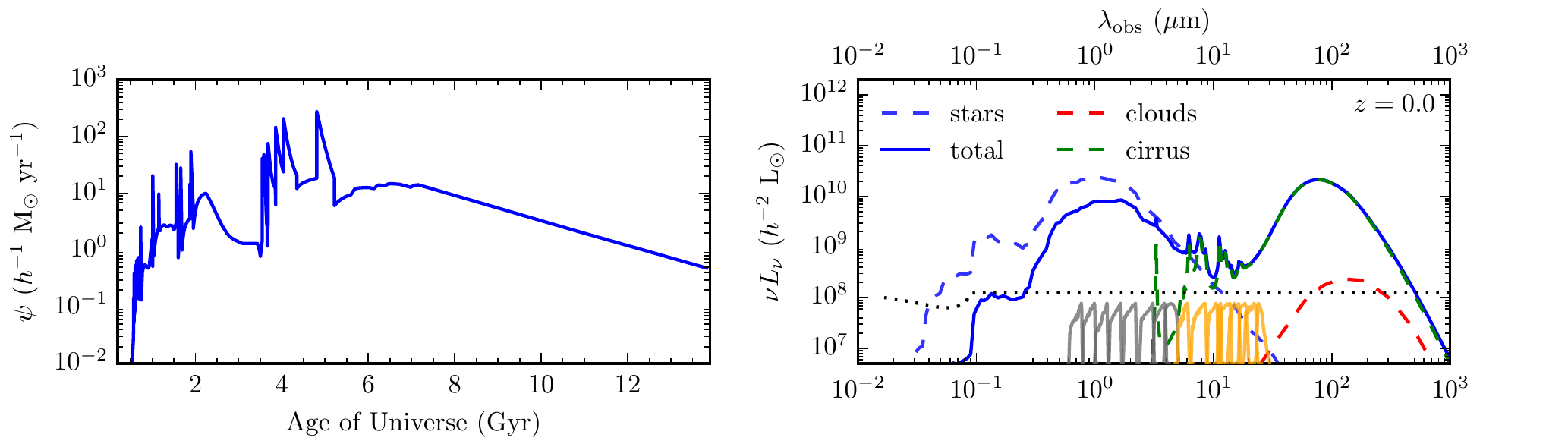}\\
\includegraphics[width=\linewidth]{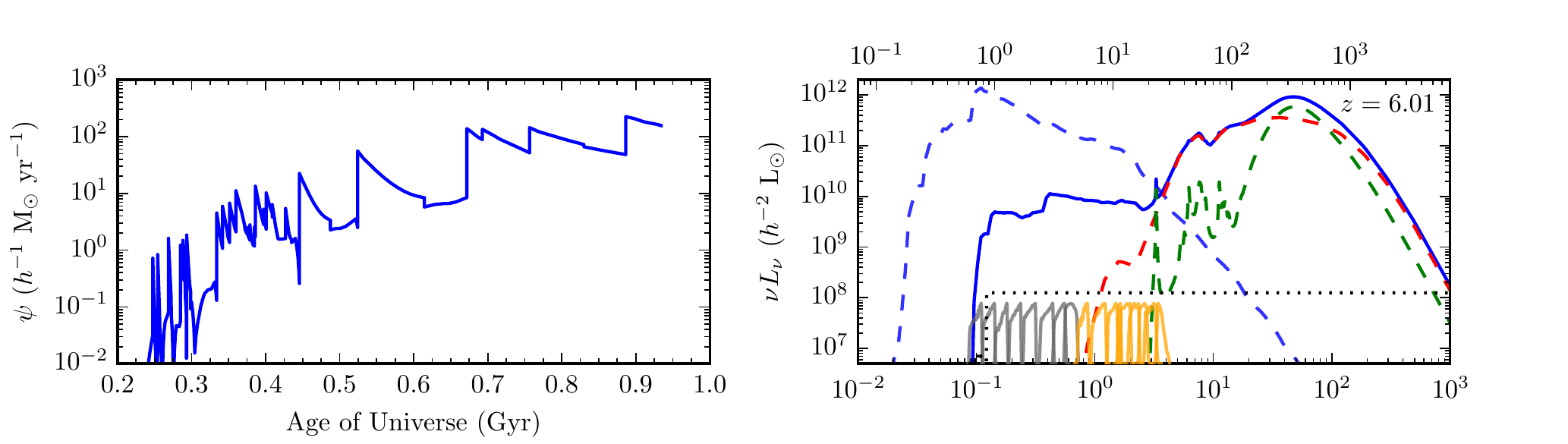}\\
\includegraphics[width=\linewidth]{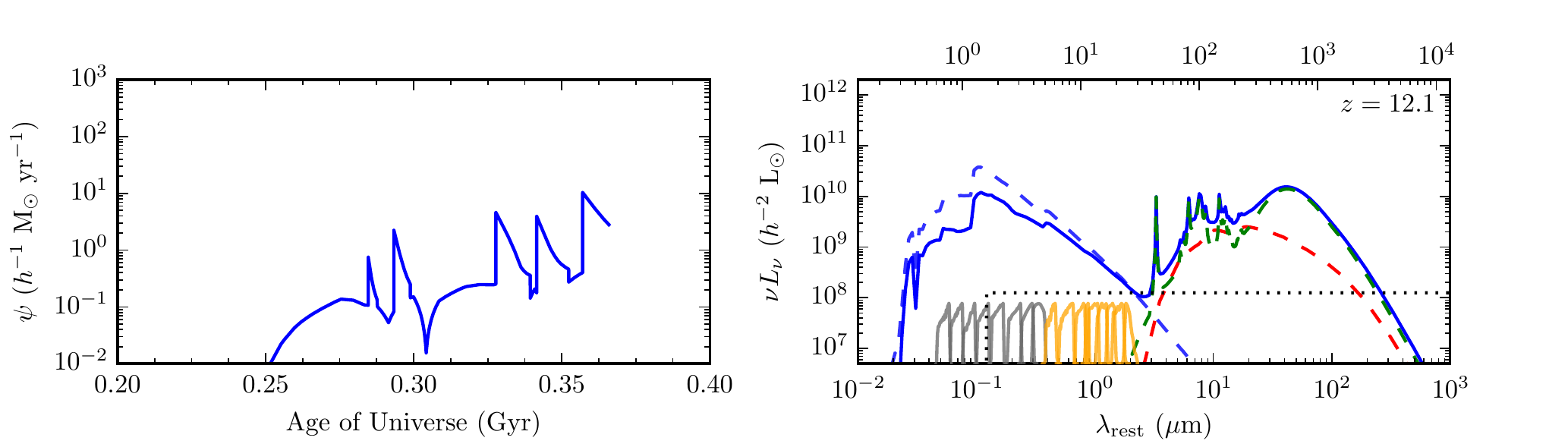}
\caption{Example galaxy star formation histories and SEDs.  Each row shows a galaxy selected at a different redshift, as indicated in the right panels.  \emph{Left panels}: star formation histories of three galaxies (in each case summed over all of the galaxy's progenitors) predicted by \galform.  Note that the range of the abscissa is different in each panel.  \emph{Right panels}: corresponding galaxy SEDs predicted by \grasil (Silva et al. \citeyear{Silva98}), plotted against rest-frame wavelength on the bottom axis and observed wavelength on the top axis.  The dashed blue line is the intrinsic stellar SED.  The solid blue line is the total galaxy SED including dust absorption and emission.  The dashed red and green lines are the dust emission for the molecular cloud and diffuse cirrus components respectively.  The \emph{JWST} filter transmission functions for NIRCam (MIRI) bands are shown in grey  (orange), in arbitrary units.  The intergalactic medium (IGM) transmission function of Meiksin (\citeyear{Meiksin05}) is shown by the dotted black line (also in arbitrary units).}
\label{fig:grasil_sed}
\end{figure*}
We use the spectrophotometric radiative transfer code \grasil\ \citep{Silva98} to compute model galaxy SEDs.  Using the star formation and metal enrichment histories, gas masses and geometrical parameters predicted by \galform, and assuming a composition and geometry for interstellar dust, \grasil\ computes the SEDs of the model galaxies, accounting for dust extinction (absorption and scattering) of radiation and its subsequent re-emission.  In this Section, we briefly describe the \grasil\ model.  For further details we refer the reader to \cite{Silva98} and \cite{Granato00}.

Here \grasil\ assumes that stars exist in a disc + bulge system, as is the case in \galform.  The disc has a radial and vertical exponential profile with scale lengths, $h_{R}$ and $h_{z}$, and the bulge is described by an analytic King model profile, $\rho\propto(r^{2}+r_{\rm c}^2)^{-3/2}$ out to a truncation radius, $r_{\rm t}$. The half-mass radii, $r_{\rm disc}$ and $r_{\rm bulge}$, are predicted by \galform.  By definition, given the assumed profiles, the bulge core radius is related to the half-mass radius by $r_{\rm c}=r_{\rm bulge}/14.6$ whilst the radial disc scale-length, $h_{\rm R}$, is related to the half-mass disc radius by $h_{R}=r_{\rm disc}/1.68$.  Star formation histories are calculated separately for the disc and bulge by \galform.  For galaxies undergoing a starburst, the burst star formation, as well as the associated gas and dust, are assumed to also be in an exponential disc but with a half-mass radius, $r_{\rm burst}=\eta r_{\rm bulge}$, rather than $r_{\rm disc}$, where $\eta$ is an adjustable parameter. The disc axial ratio, $h_{z}/h_{R}$, is a parameter of the \grasil\ model; for starburst galaxies, the axial ratio of the burst is allowed to be different from that of discs in quiescent galaxies.

The gas and dust exist in an exponential disc, with the same radial scale-length as the disc stars but in general with a different scale-height, so $h_{z}\mathrm{(dust)}/h_{z}\mathrm{(stars)}$ is an adjustable parameter.  The gas and dust are assumed to exist in two components: (i) giant molecular clouds in which stars form, escaping on some time scale, $t_{\rm esc}$, and (ii) a diffuse cirrus ISM.  The total gas mass, $M_{\rm cold}$, and metallicity, $Z_{\rm cold}$, are calculated by \galform.  The fraction of gas in molecular clouds is determined by the parameter $f_{\rm cloud}$.  The cloud mass, $m_{\rm cloud}$, and radius, $r_{\rm cloud}$, are also parameters, though the results of the model depend only on the ratio, $m_{\rm cloud}/r_{\rm cloud}^2$, which determines (together with the gas metallicity) the optical depth of the clouds.  

The dust is assumed to consist of a mixture of graphite and silicate grains and polycyclic aromatic hydrocarbons (PAHs), each with a distribution of grain sizes.  The grain mix and size distribution were determined by Silva et al. so that the extinction and emissivity properties of the local ISM are reproduced using the optical properties of the dust grains tabulated by \cite{DraineLee84}.  At long wavelengths ($\lambda>30$~$\mu$m) this results in a dust opacity that approximates $\kappa_{\rm d}\propto\lambda^{-2}$.  However, in galaxies undergoing a starburst this is modified (for $\lambda>100$~$\mu$m) such that $\kappa_{\rm d}\propto\lambda^{-\beta_{\rm b}}$, where $\beta_{\rm b}$ is treated as an adjustable parameter.  Laboratory measurements suggest that values in the range $\beta_{\rm b}=1.5-2$ are acceptable (Agladze et al. \citeyear{Agladze96}).  Here a value of $\beta_{\rm b}=1.5$ is adopted \citep{Lacey16}.  The total dust mass in a galaxy is proportional to the cold gas mass and metallicity, both of which are predicted by \galform.  

The adopted values of adjustable \grasil parameters are summarised in Table~\ref{table:grasil_params}.   
\begin{table}
\centering 
\caption{Adopted values for adjustable parameters in \grasil.  See the text in Section~\ref{subsec:grasil} for their definitions.}
\begin{tabular}{cc}\hline
Parameter & Value\\\hline
$h_{z}/h_{R}$ (disc) & $0.1$ \\
$h_{z}/h_{R}$ (burst) & $0.5$ \\
$h_{z}\mathrm{(dust)}/h_{z}\mathrm{(stars)}$& $1$ \\
$\eta$ & $1.0$ \\
$f_{\rm cloud}$& $0.5$ \\
$m_{\rm cloud}/r_{\rm cloud}^2$ & $10^{6}$~M$_{\sun}/(16$~pc$)^2$ \\
$t_{\rm esc}$ & $1$~Myr \\
$\beta_{\rm b}$ & $1.5$ \\\hline
\end{tabular}
\label{table:grasil_params}
\end{table} For the parameters which are analogous to those in the dust model used by \cite{Lacey16}: $f_{\rm cloud}$, $m_{\rm cloud}/r_{\rm cloud}^{2}$, $t_{\rm esc}$ and $\beta_{\rm b}$, we use the values chosen by Lacey et al.  For other parameters specific to the \grasil model, we use the values chosen by by Baugh et al. (\citeyear{Baugh05}, see also Lacey et al. \citeyear{Lacey08}, Swinbank et al. \citeyear{Swinbank08} and Lacey et al. \citeyear{Lacey11}), which was the last time a published version of \galform\ was coupled with \grasil in the manner presented here. 

The luminosities of the stellar components are calculated assuming the \cite{Maraston05} evolutionary population synthesis model, as is done in \cite{Lacey16}.  \grasil then calculates the radiative transfer of the stellar radiation through the interstellar dust.  For molecular clouds, a full radiative transfer calculation is performed.  For the diffuse cirrus the effects of scattering are included approximately by using an effective optical depth for the absorption $\tau_{\rm abs,eff}=[\tau_{\rm abs}(\tau_{\rm abs}+\tau_{\rm scat})]^{1/2}$.  The dust-attenuated stellar radiation field can be calculated at any point inside or outside the galaxy.  \grasil then computes the final galaxy SED by calculating the absorption of stellar radiation, thermal balance and the re-emission of radiation for each grain species and size at every point in the galaxy.

Examples of predicted star formation histories and the resulting galaxy UV-to-mm SEDs computed by \grasil are shown in Fig.~\ref{fig:grasil_sed}.  One can see that the star formation histories are extremely `bursty' at early times when the Universe is a few Gyr old.  Significant dust extinction and re-emission is evident for each of the galaxy SEDs shown.  There are also a number of interesting features in the galaxy SEDs.  These include: (i) Lyman-continuum breaks in the galaxy SEDs at $912$~{\AA}; (ii) a prominent $4000$~{\AA} break for the $z=0$ galaxy, indicative of an old stellar population (which would be expected from the smoothly declining star formation history of this galaxy); (iii) dust emission approximating a modified blackbody that peaks at $\lambda_{\rm rest}\approx100$~$\mu$m, indicative of cold ($\sim30$~K) dust, though the peak of the emission shifts to shorter wavelengths with increasing redshift suggesting hotter dust, and (iv) PAH emission lines in the cirrus dust at $\lambda_{\rm rest}=3.3$, $6.2$, $7.7$, $8.6$, and $11.3$~$\mu$m.

Once an SED has been computed, luminosities in specified bands are calculated by convolving the SED (redshifted into the observer frame) with the filter transmission of interest.  We use the \cite{Meiksin05} prescription for attenuation of radiation in the IGM due to neutral hydrogen, also shown in Fig.~\ref{fig:grasil_sed}.          
\subsection{Coupling \galform\ and \grasil} 
\label{subsec:coupling}
Here we briefly describe how the \galform\ and \grasil models are used in conjunction.  For further details, we refer the reader to \cite{Granato00}.

Due to the computational expense of running \grasil ({$\sim3-5$}~CPU mins per galaxy) it is not feasible to compute an SED for each galaxy in the simulation volume, as has been discussed in previous studies \citep[e.g.][]{Granato00,Almeida10,Lacey11}.  However, for the purposes of constructing luminosity functions, it is possible to circumvent this by running \grasil on a sample of galaxies, from which the luminosity function can be constructed if the galaxies in question are weighted appropriately.  We choose to sample galaxies according to their stellar mass such that $\sim10^3$ galaxies per dex of stellar mass are sampled.  We use a lower mass limit of $10^6$~{\hMsol}, which we choose so that any artificial features it introduces into our predicted luminosity functions (see Section~\ref{subsec:lfs}) are at fainter luminosities than those investigated here.  This represents a factor of $\sim10$ increase over the number of galaxies sampled by \cite{Granato00}.

The procedure that we use to construct luminosity functions in a given band at each output redshift is as follows: (i) run \galform\ to the redshift of interest; (ii) create a subsample of galaxies; (iii) re-run \galform\ to output the star formation and metal enrichment history for each of the sampled galaxies; (iv) run \grasil on each of the sampled galaxies to produce a predicted SED; (v) convolve the output SED with the relevant broadband filter response and IGM attenuation curve \citep{Meiksin05} and (vi) construct the galaxy luminosity function using the weights from the initial sampling and luminosities from the previous step.

We have made a number of improvements to steps (iii) to (v) above, which allow us to run \grasil for samples of $\sim10^{5}$ galaxies for each model, spread over $25$ output redshifts from $z=16$ to $z=0$.  For each model, this takes $\sim7\times10^3$~CPU~hours, approximately $95$~per~cent of which is spent by \grasil, with the remaining time being taken by \galform\ to calculate the necessary star formation histories.
\section{Results}
\label{sec:Results}
In this Section, we present our main results.  In Section~\ref{subsec:L16_at_hi_z} we present predictions for the evolution of physical properties of the galaxy population as well as a comparison of our predictions with available high-redshift ($z\gtrsim7$) observational data.  In Section~\ref{subsec:lfs} we present the predicted evolution of the galaxy luminosity function for the NIRCam--F200W and MIRI--F560W filters.  We make such predictions for each NIRCam and MIRI broadband filter but only show these two in this paper for brevity; results for other filters will be made available online.  In Section~\ref{subsec:ncts_dndz} we present predictions for galaxy number counts and redshift distributions (for a $10^{4}$~s exposure) observable by \emph{JWST} in each NIRCam and MIRI band; we also show predictions for the redshift distributions of galaxies observable with longer ($10^{5}$ and $10^{6}$~s) exposures.  Finally, in Section~\ref{subsec:JWST_sizes} we present predictions for the angular sizes of galaxies for the NIRCam--F200W and MIRI--F560W filters, again we make such predictions for all NIRCam filters but show only these two here for brevity.  Throughout we show predictions for our fiducial model `lc16' and the variant `lc16.EvolFB' that adopts the evolving feedback model presented in \cite{HouJun16} and is discussed in Section~\ref{subsec:EvolFB}.  The dependence of our high-redshift predictions on some assumptions made in the model is discussed briefly in Appendix~\ref{sec:discussion}.   
\subsection{The Lacey et al. (\citeyear{Lacey16}) model at high redshift}
\label{subsec:L16_at_hi_z}
\begin{figure*}
\includegraphics[width = \linewidth]{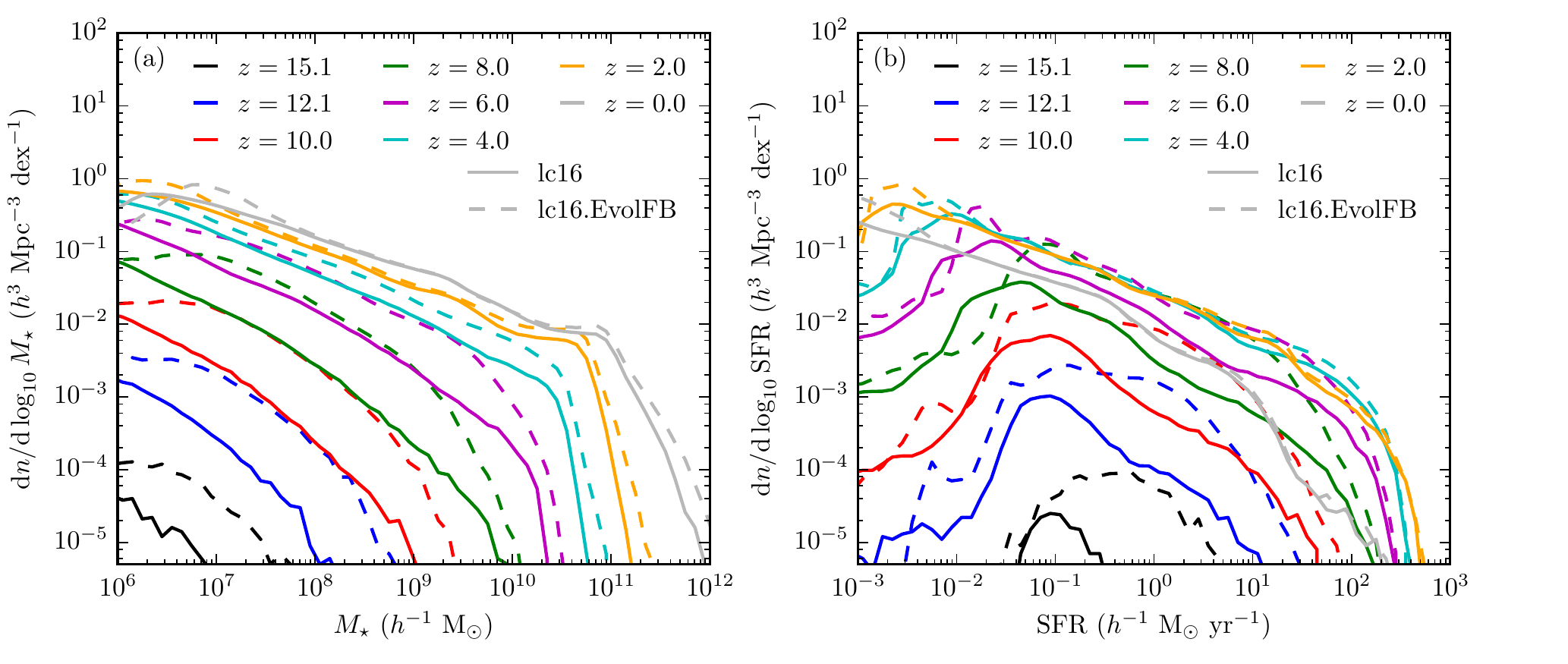}\\
\includegraphics[width = \linewidth]{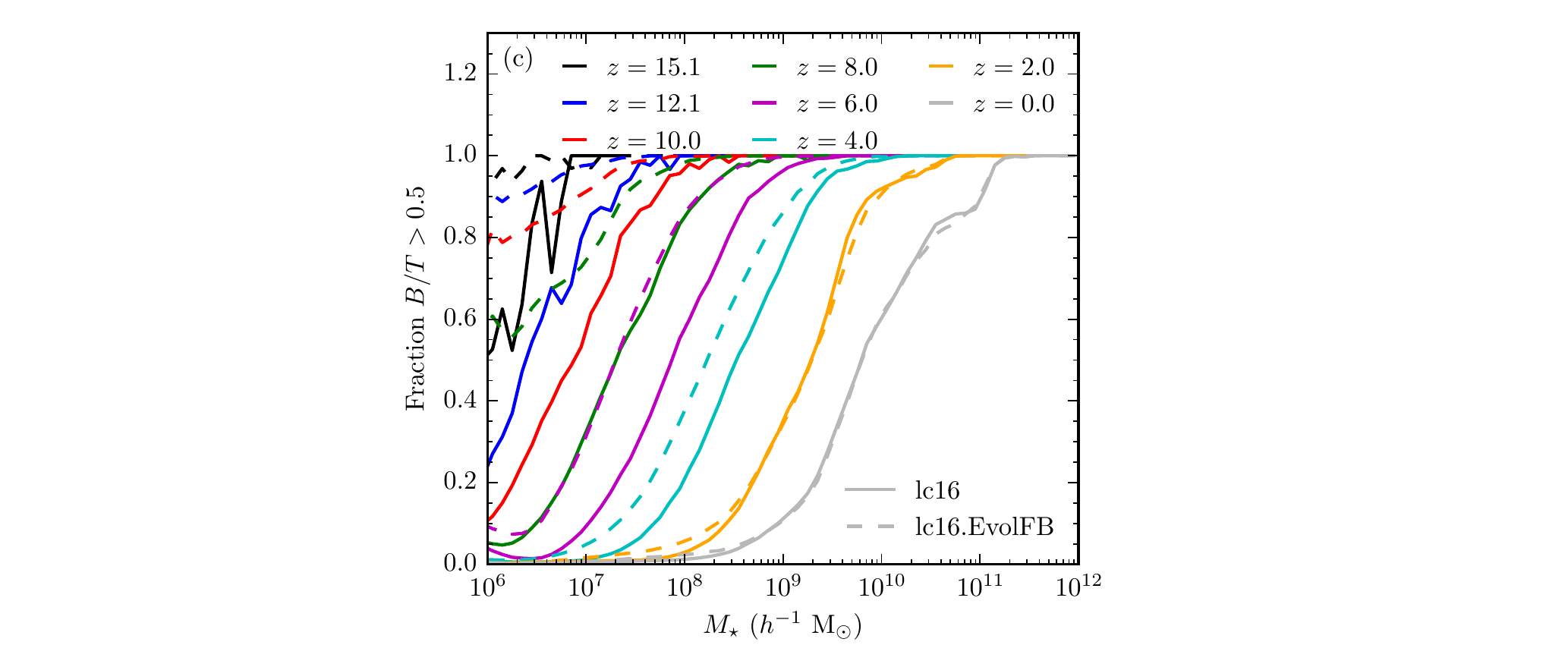}
\caption{Predicted evolution of physical galaxy properties from $z=15.1$ to $z=0$.  \emph{Panel (a)}: the galaxy stellar mass function.  \emph{Panel (b)}: the star formation rate function for galaxies with $M_{\star}>10^{6}$~{\hMsol}.  \emph{Panel (c)}: the fraction of bulge-dominated (bulge-to-total stellar mass ratios, $B/T>0.5$) galaxies as a function of stellar mass.  In each panel, the colour of the line indicates the redshift as shown in the legend.  The solid lines are predictions from the fiducial model whereas the dashed lines are predictions from the evolving feedback variant.}
\label{fig:physical_props}
\end{figure*}
\begin{figure}
\centering
\includegraphics[trim = 0 0 4.135in 0,clip = True,width= \linewidth]{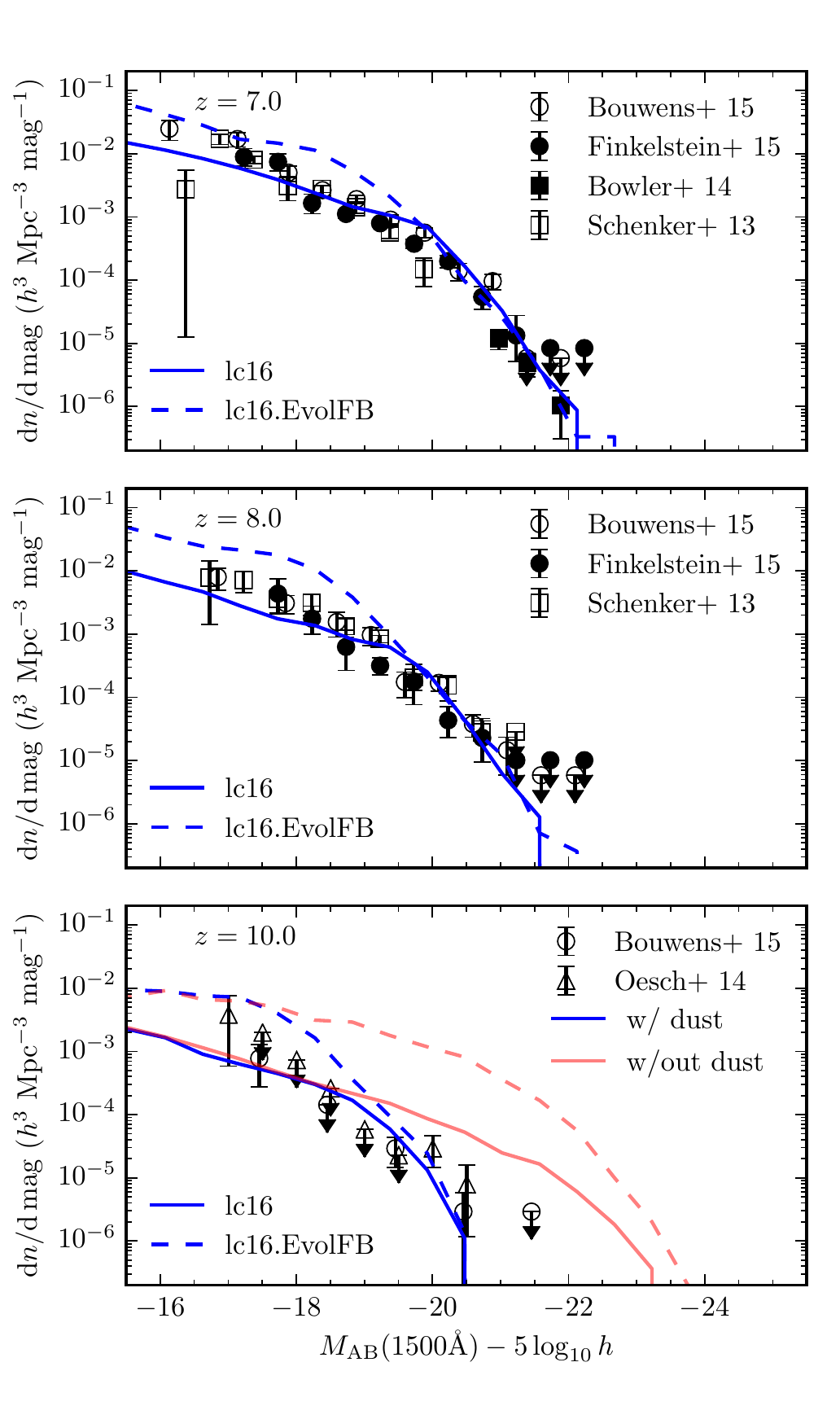}
\caption{The predicted rest-frame far-UV ($1500$~{\AA}) luminosity functions for $z=7-10$ for the fiducial model (solid blue line) and the evolving feedback variant (dashed blue line).  The redshift is indicated in each panel.  Observational data are from Bouwens et al. (\citeyear{Bouwens15}, open circles), Finkelstein et al. (\citeyear{Finkelstein15}, filled circles), Bowler et al. (\citeyear{Bowler14}, filled squares), Schenker et al. (\citeyear{Schenker13}, open squares) and Oesch et al. (\citeyear{Oesch14}, open triangles) as indicated in the legend.  In the bottom panel, the red lines show the model predictions without dust extinction.}
\label{fig:UVlfs}
\end{figure}
\begin{figure*}
\centering
\includegraphics[width = \linewidth]{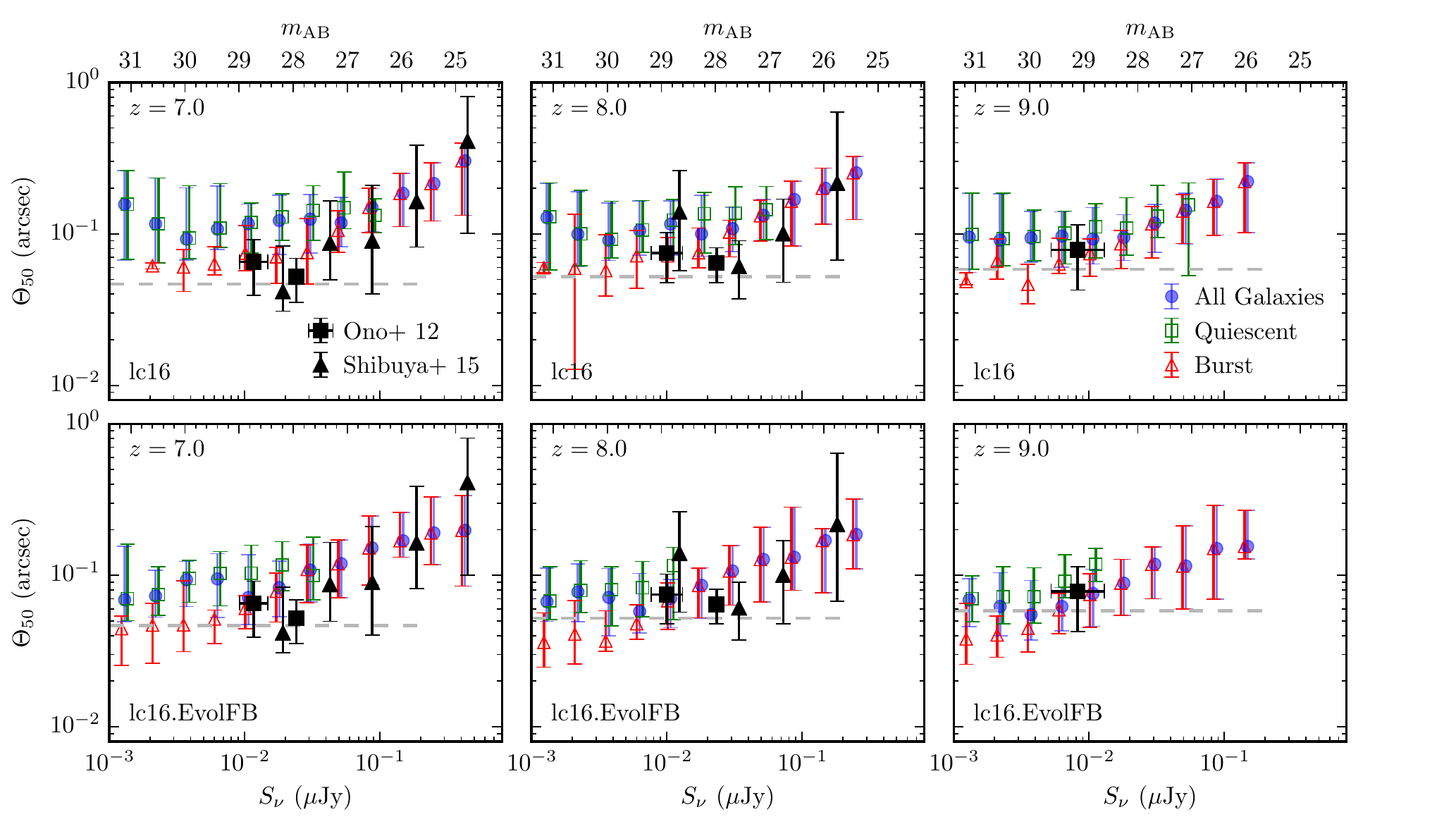}
\caption{Predicted rest-frame far-UV ($1500$~{\AA}) galaxy projected half-light radii for $z=7-9$, as a function of galaxy flux, $S_{\nu}$.  The redshift is indicated in each panel.  The top row shows predictions from the fiducial model, whereas the bottom row shows predictions from the evolving feedback variant.  Blue filled circles indicate the median size for all galaxies at a given flux, with the error bars indicating the $16-84$ percentile range.  The open green squares and red triangles indicate this for quiescent and starburst galaxies respectively.  Observational data are from Ono et al. (\citeyear{Ono13}, black filled squares) and Shibuya et al. (\citeyear{Shibuya15}, black filled triangles).  For reference, the horizontal dashed line in each panel indicates the diffraction limit for \emph{JWST} for a fixed rest-frame wavelength of $1500$~{\AA}, assuming a $6.5$~m diameter mirror.}
\label{fig:sizes}
\end{figure*}
In this Section, we present model predictions for the evolution of some physical properties of the galaxy population and compare our predictions at $z\gtrsim7$ to available observational data.  In Fig.~\ref{fig:physical_props} we show predictions of the fiducial and evolving feedback variant models for the evolution of: (a) the galaxy stellar mass function; (b) the galaxy star formation rate function (for $M_{\star}>10^{6}$~{\hMsol} galaxies); and (c) the fraction of bulge-dominated (i.e. with bulge-to-total stellar mass ratios of $B/T>0.5$) galaxies as a function of stellar mass, from $z=15.1$ to $z=0$.  

The stellar mass function [Fig.~\ref{fig:physical_props}~(a)] evolves rapidly at $z\gtrsim2$ in both models.  At lower redshifts, further evolution is predominantly at the high-mass end.  It is easily seen that (for $z\gtrsim2$) the evolving feedback model results in both more massive galaxies and a greater abundance of galaxies at a given stellar mass (for $M_{\star}\gtrsim10^{6}$~{\hMsol}, as galaxies with a lower stellar mass are not included in our \grasil sampling) by factors of up to $\sim10$.  For $z<4$, the normalisation of the supernova feedback strength is the same in both models and the differences between their stellar mass functions begin to disappear.  At the low mass end ($M_{\star}\lesssim10^{8}$~{\hMsol}), however,  the break in the power law for the mass loading factor (at $V_{\rm thresh}=50$~km~s$^{-1}$) in the evolving feedback model results in a greater abundance of galaxies at these stellar masses than in the fiducial model. At the high mass end ($M_{\star}\gtrsim10^{11}$~{\hMsol}), an increase in stellar mass at low redshift due to the reduced feedback strength at higher redshift is apparent.

The distributions of star formation rates [Fig.~\ref{fig:physical_props}~(b)] tell a similar story.  For $z<4$ the distributions predicted by both models are essentially identical, except at low star formation rates (SFRs$\,\lesssim10^{-2}$~$h^{-1}$~M$_{\sun}$~yr$^{-1}$) where the break in the evolving feedback model results in this model having a greater abundance of galaxies.  At higher redshifts $z>4$ the differences in the star formation rate distributions are greater due to the different normalisations of feedback, with the evolving feedback variant having significantly more galaxies with SFRs$\,\gtrsim3\times10^{-2}$~$h^{-1}$~M$_{\sun}$~yr$^{-1}$.  The apparent peak seen in each SFR distribution is mostly due to the imposed stellar mass limit of $10^{6}$~{\hMsol}, if lower stellar mass galaxies were included it would shift to lower star formation rates according to the (approximately) constant relation between specific star formation rate and stellar mass predicted by the model \citep[e.g.][]{Mitchell14,Cowley16SEDs}.

Fig.~\ref{fig:physical_props}~(c) shows the evolution in the fraction of galaxies with a bulge-to-total stellar mass ratio of $B/T>0.5$, as a function of total stellar mass.  In \galform, bulges are created by a dynamical process, either a galaxy merger or a disc instability.  The transition from a disc-dominated to a bulge-dominated galaxy population is relatively sharp, occurring over roughly one dex in stellar mass in most cases.  In the evolving feedback model, this transition generally occurs at lower stellar masses.  At higher redshifts (and thus lower stellar masses), the shape of this relationship is different for the evolving feedback variant, which predicts a much smoother transition.  We caution against over-interpreting the predicted $B/T$ as a proxy for the morphological type.  The instabilities that create bulges in \galform\ do not necessarily create slowly rotating bulges, and so defining bulges as slow rotators would give different results to those presented here. 

Having established some predicted physical properties of galaxies in the two models, we now compare predictions of the models to observational data at $z\gtrsim7$.  We note that none of the observational data considered here were used to calibrate model parameters [Lacey et al. (\citeyear{Lacey16}) only considered rest-frame far-UV luminosity functions at $z\lesssim6$ in their model calibration].

We compare the predictions of the models for the evolution of the rest-frame far-UV luminosity function to observational data over the redshift interval $7\lesssim z\lesssim10$ in Fig.~\ref{fig:UVlfs}.  We can see that both models provide reasonable agreement with the observed data, and appear to `bracket' the data for $M_{\rm AB}(1500~\AA)-5\log_{10}h\gtrsim-18$.  However, at brighter magnitudes, the predictions of the two models converge.  This is due to dust extinction becoming the limiting factor in a galaxy's intrinsic brightness at far-UV wavelengths.  To illustrate this, we show the predictions of the two models, without dust attenuation, in the $z=10$ panel.  These predictions resemble the star formation rate distributions in Fig.~\ref{fig:physical_props}~(b), as the star formation rate of a galaxy is essentially traced by the rest-frame far-UV.

Finally, we compare predictions for the angular sizes of galaxies to observational data in the redshift range $7\lesssim z\lesssim9$ in Fig~\ref{fig:sizes}.  The stellar component of the model galaxies is assumed to be a composite system, consisting of an exponential disc and a bulge with a projected $r^{1/4}$ density profile \citep{Cole00}.  We compute the half-light radii for our model galaxies by weighting the density profile of each component by their predicted rest-frame far-UV ($1500$~{\AA}) luminosity, dividing the half-light radii of the disc by a factor of $1.34$ to account for inclination effects \citep{Lacey16}, and interpolating to find the half-light radius of the composite system.  We then bin the galaxies according to their flux, $S_{\nu}$.  The symbols in Fig.~\ref{fig:sizes} show the median size in each flux bin, with the error bars representing the $16-84$~percentile scatter in each bin.  We show this for the whole galaxy population and also for starburst and quiescent galaxies.  The differences between the predictions of the two models are small and they both show reasonable agreement with data from \cite{Ono13} and \cite{Shibuya15}, who use {\sc{GALFIT}\xspace} (Peng et al. \citeyear{Peng02}) to derive sizes from \emph{Hubble Space Telescope} imaging.  For the Ono et al. data we present their stacked image results.  For the Shibuya et al. data we bin their sizes for individual galaxies into bins of $1$~mag width and present the median size in each bin.  The error bars presented represent the $16-84$~percentile scatter of sizes within these bins.  For reference, we also show the diffraction limit of \emph{JWST}.  The models predict that \emph{JWST} should be able to resolve most galaxies in the rest-frame far-UV at these redshifts.

In summary, the predictions of both models show good agreement with the evolution of the rest-frame far-UV ($1500$~{\AA}) luminosity function and observed galaxy sizes at high redshift ($z\gtrsim7$).  We re-iterate that these high-redshift data were not considered when calibrating the model.
\subsection{Luminosity functions observable with \emph{JWST}}
\label{subsec:lfs}
\begin{figure*}
\includegraphics[width = \linewidth]{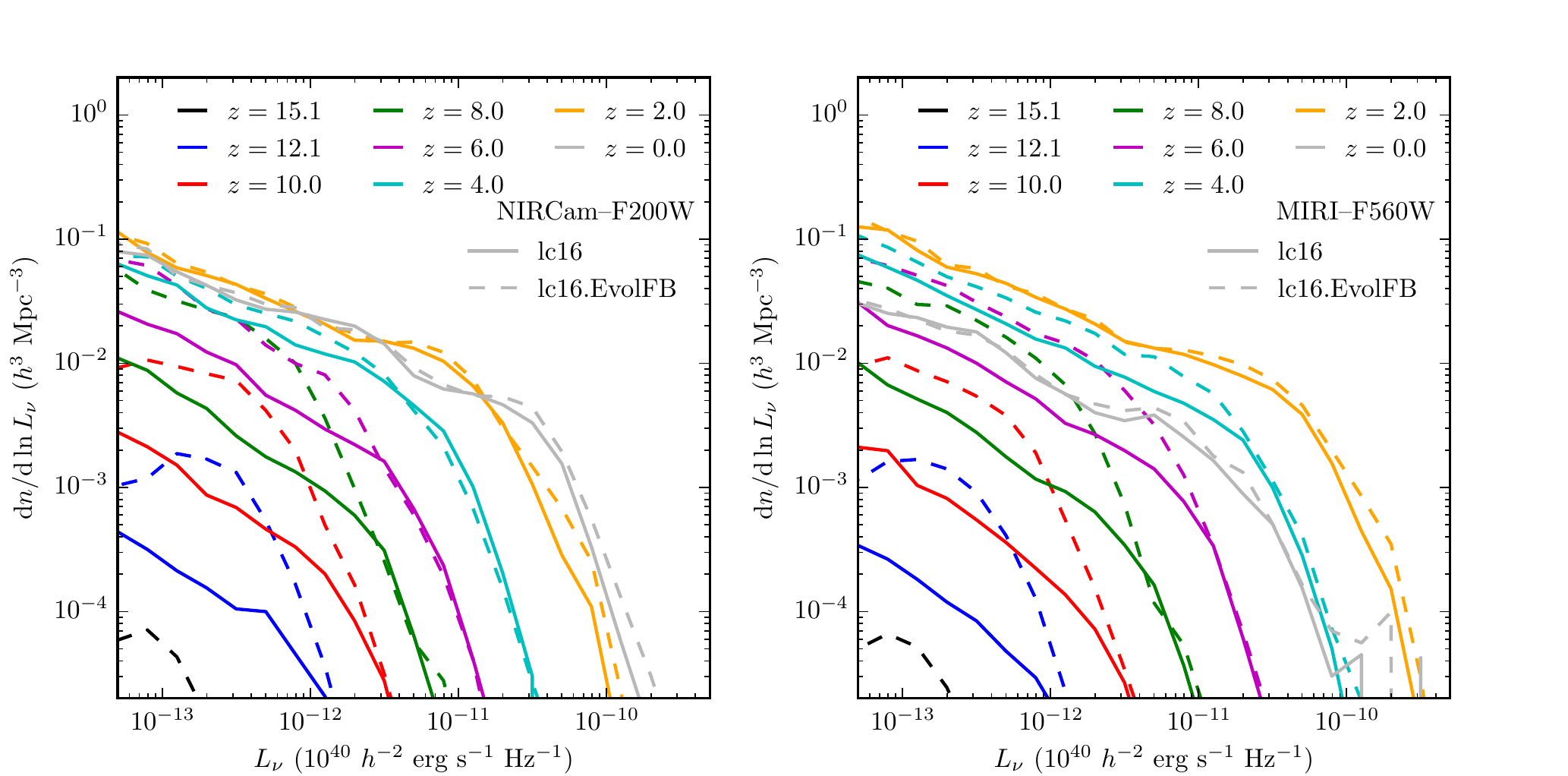}\\
\vspace*{0.2cm}
\includegraphics[width = \linewidth]{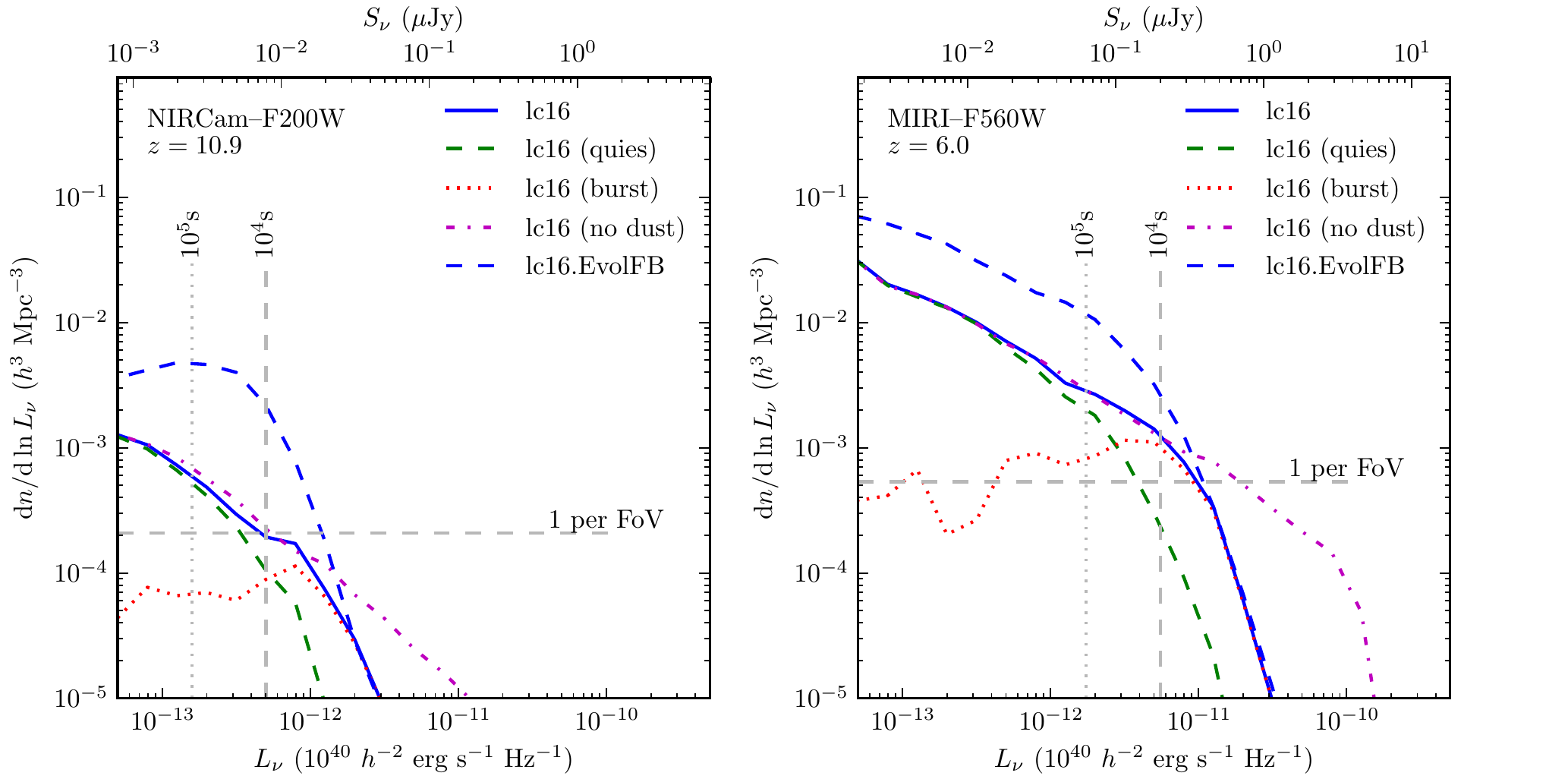}
\caption{\emph{Top panels}: predicted evolution from $z=15.1$ to $z=0.0$ of the luminosity function in the NIRCam--F200W (left panel) and MIRI--F560W (right panel) bands (in the observer-frame).  The colour indicates the redshift as shown in the legend.  The solid lines show predictions from the fiducial model, whereas the dashed lines show predictions of the evolving feedback variant.  \emph{Bottom panels}: a breakdown of the predicted luminosity functions for NIRCam--F200W at $z=10.9$ (left panel) and MIRI--F560W at $z=6.0$ (right panel).  The solid blue lines show the predictions of the fiducial model and the dashed green and dotted red lines show the contribution to this from quiescent and starburst galaxies respectively.  The predictions of the fiducial model excluding dust absorption are shown by the dash-dotted magenta lines.  The dashed blue line is the prediction from the evolving feedback model.  For reference, the horizontal dashed lines indicate the number density at which there is one object per \emph{JWST} field of view at that redshift and the vertical dashed and dotted lines indicate the \emph{JWST} sensitivity limits for that filter for a $10^{4}$ and $10^{5}$~s exposure, as labelled.}
\label{fig:JWST_lfs}
\end{figure*}
\begin{table}
\centering
\caption{Adopted sensitivities for \emph{JWST} filters based on $10\sigma$ point source and $10^{4}$~s exposure.}
\begin{tabular}{llll}\hline
Instrument & Filter & $\lambda_{\rm eff}$ ($\muup$m) & Sensitivity ($\muup$Jy) \\ \hline
\rule{0pt}{2.5ex}NIRCam & F070W & $0.70$ & $20.9\times10^{-3}$\\
&F090W & $0.90$&$13.1\times10^{-3}$\\
&F115W & $1.15$&$11.8\times10^{-3}$\\
&F150W & $1.50$&$9.6\times10^{-3}$\\
&F200W & $2.00$&$7.9\times10^{-3}$\\
&F277W & $2.77$&$11.5\times10^{-3}$\\
&F356W & $3.56$&$11.1\times10^{-3}$\\
&F444W & $4.44$&$17.6\times10^{-3}$\\\hline
MIRI&F560W&$5.6$&$0.2$\\
&F770W&$7.7$&$0.28$\\
&F1000W&$10.0$&$0.7$\\
&F1130W&$11.3$&$1.7$\\
&F1280W&$12.8$&$1.4$\\
&F1500W&$15.0$&$1.8$\\
&F1800W&$18.0$&$4.3$\\
&F2100W&$21.0$&$8.6$\\
&F2550W&$25.5$&$28.0$\\\hline
\multicolumn{4}{p{0.85\columnwidth}}{\raggedright{\textbf{Note:} Adapted from \url{https://jwst.stsci.edu/files/live/sites/jwst/files/home/science\%20planning/Technical\%20documents/JWST-PocketBooklet_January17.pdf}}}\\\hline
\end{tabular}
\label{table:JWST_sensitivities}
\end{table}
\begin{table}
\centering
\caption{Adopted \emph{JWST} instrument fields of view (FoV).}
\begin{tabular}{lc}\hline
\rule{0pt}{2.5ex}Instrument & FoV (arcmin$^2$)\\\hline
NIRCam & $2\times2.2\times2.2$\\
MIRI & $1.23\times1.88$\\\hline
\multicolumn{2}{p{0.85\columnwidth}}{\raggedright{\textbf{Note:} From \url{https://jwst.stsci.edu/files/live/sites/jwst/files/home/science\%20planning/Technical\%20documents/JWST-PocketBooklet_January17.pdf}}}\\\hline
\end{tabular}
\label{table:JWST_FoV}
\end{table}
In this Section we present predictions for the evolution of the galaxy luminosity function in the \emph{JWST} NIRCam and MIRI bands.  These are listed in Table~\ref{table:JWST_sensitivities}, with their sensitivities (for a $10^{4}$~s exposure), and the field of view (FoV) for each instrument is shown in Table~\ref{table:JWST_FoV}.  In Fig.~\ref{fig:JWST_lfs} we show the predicted luminosity functions for the NIRCam--F200W and MIRI--F560W bands.  We make such predictions for all broadband NIRCam and MIRI filters, but show only these two here for brevity.  The predictions for other filters will be made available online.  

In the top panels of Fig.~\ref{fig:JWST_lfs} we can see that at high redshifts the difference between the two models is similar to that seen in Fig.~\ref{fig:UVlfs}, and that the models predict similar luminosity functions for $z<4$, when the normalisation of the feedback strength is the same in both models.  

In the bottom panels, we show the predicted luminosity function at $z=11$ for NIRCam--F200W (bottom left panel), and at $z=6$ for MIRI--F560W (bottom right panel).  We choose these values as they are the redshifts at which we predict \emph{JWST} will see $\sim1$ object per field of view (FoV) for a $10^{4}$~s exposure, as is discussed below.  Here we show the contribution to the luminosity function predicted by the fiducial model from quiescent and starburst galaxies.  We can see that the bright end of the luminosity function is dominated by galaxies undergoing a burst of star formation.  As mentioned earlier, the definition of starburst here refers to a dynamical process, either a galaxy merger or disc instability, triggering a period of enhanced star formation.  In this case, the majority of the bursts are triggered by disc instabilities, as mergers appear to be inefficient at boosting the specific star formation rates of galaxies in this model, as is also discussed in \cite{Cowley16SEDs}. We also show predictions of the fiducial model without dust and can see that the bright end of the luminosity functions at these redshifts is composed of heavily dust-attenuated objects.  We, therefore, expect such observations to provide a further constraint on the way dust absorption is accounted for in galaxy formation models.

For reference, we have also shown the sensitivity limits of the filters based on $10^4$ and $10^5$~s exposures as the vertical dashed and dotted lines respectively.  Our adopted sensitivities for a $10^4$~s exposure are summarised in Table~\ref{table:JWST_sensitivities}.  We derive sensitivities for other exposures assuming they scale as $t^{-1/2}$.

In conjunction, we also show the abundance at which the instrument will see one object per FoV per unit redshift at this redshift.  Our adopted fields of view are summarised in Table~\ref{table:JWST_FoV}.  Objects that are in the upper right quadrant of each plot would be observable with a $10^{4}$~s exposure in a single FoV.  Therefore, the fiducial model predicts that $\sim1$ object will be observable at $z=11$ by NIRCam--F200W, and $\sim2$ will be observable at $z=6$ by MIRI--F560W.  We recognise that single FoV observations will be sensitive to field-to-field variance. We hope to make direct predictions for the field-to-field variance by creating lightcone catalogues from our simulation in a future work.
\subsection{Galaxy number counts and redshift distributions observable with \emph{JWST}}
\label{subsec:ncts_dndz}
\begin{figure*}
\includegraphics[width = \linewidth]{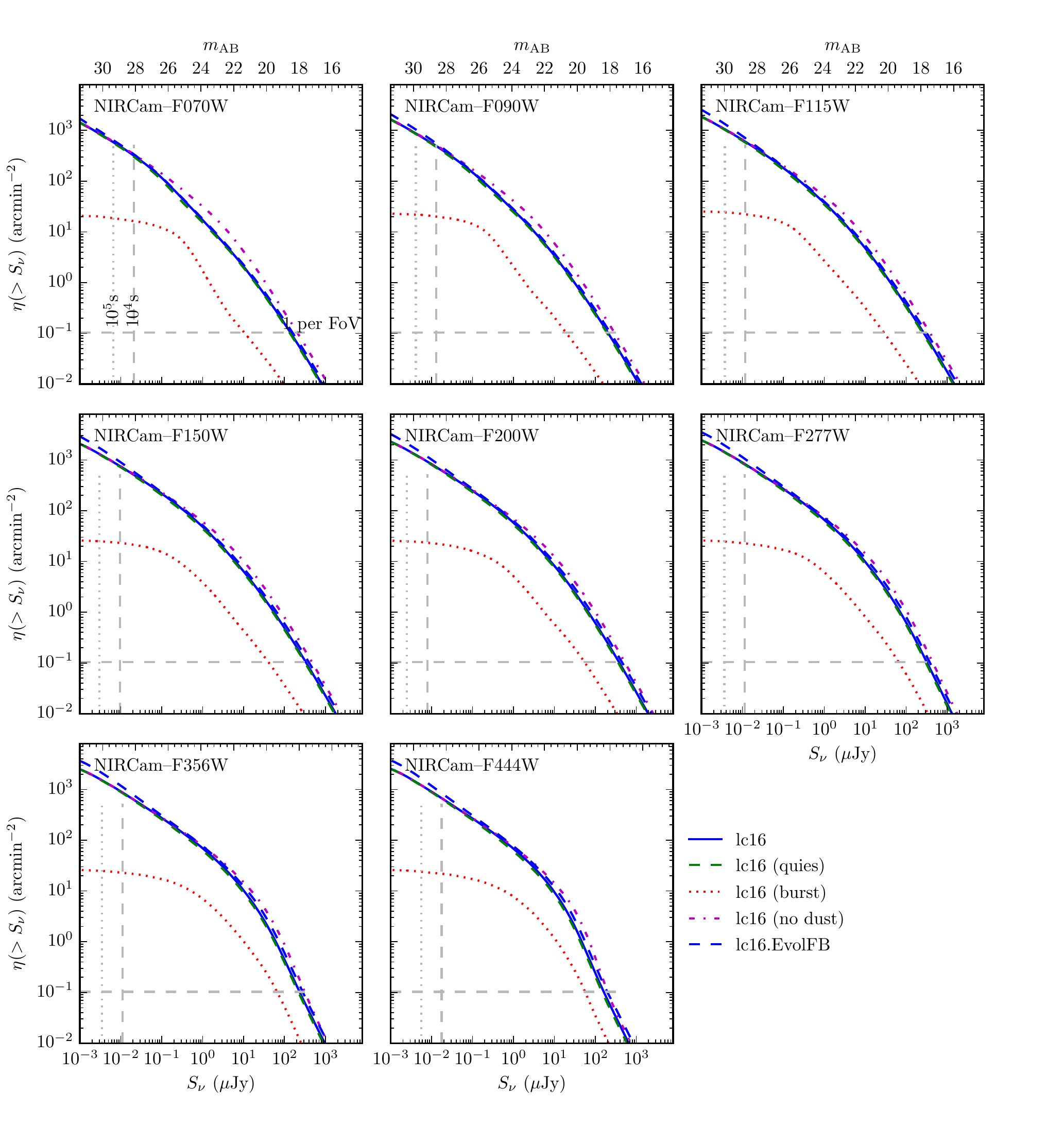}
\caption{Predicted cumulative galaxy number counts in the NIRCam bands.  The name of the band is indicated in each panel.  The solid blue lines show the predictions of the fiducial model and the dashed green and dotted red lines show the contribution to this from quiescent and starburst galaxies respectively.  The predictions of the fiducial model excluding dust absorption are shown by the dash-dotted magenta lines.  The dashed blue lines show the predictions from the evolving feedback variant.  For reference, the horizontal dashed lines indicate the number density at which there is one object per field of view and the vertical dashed and dotted lines indicate the sensitivity limits for that filter for a $10^{4}$ and $10^{5}$~s exposure respectively.}
\label{fig:cncts_NIRCam}
\end{figure*}
\begin{figure*}
\includegraphics[width = \linewidth]{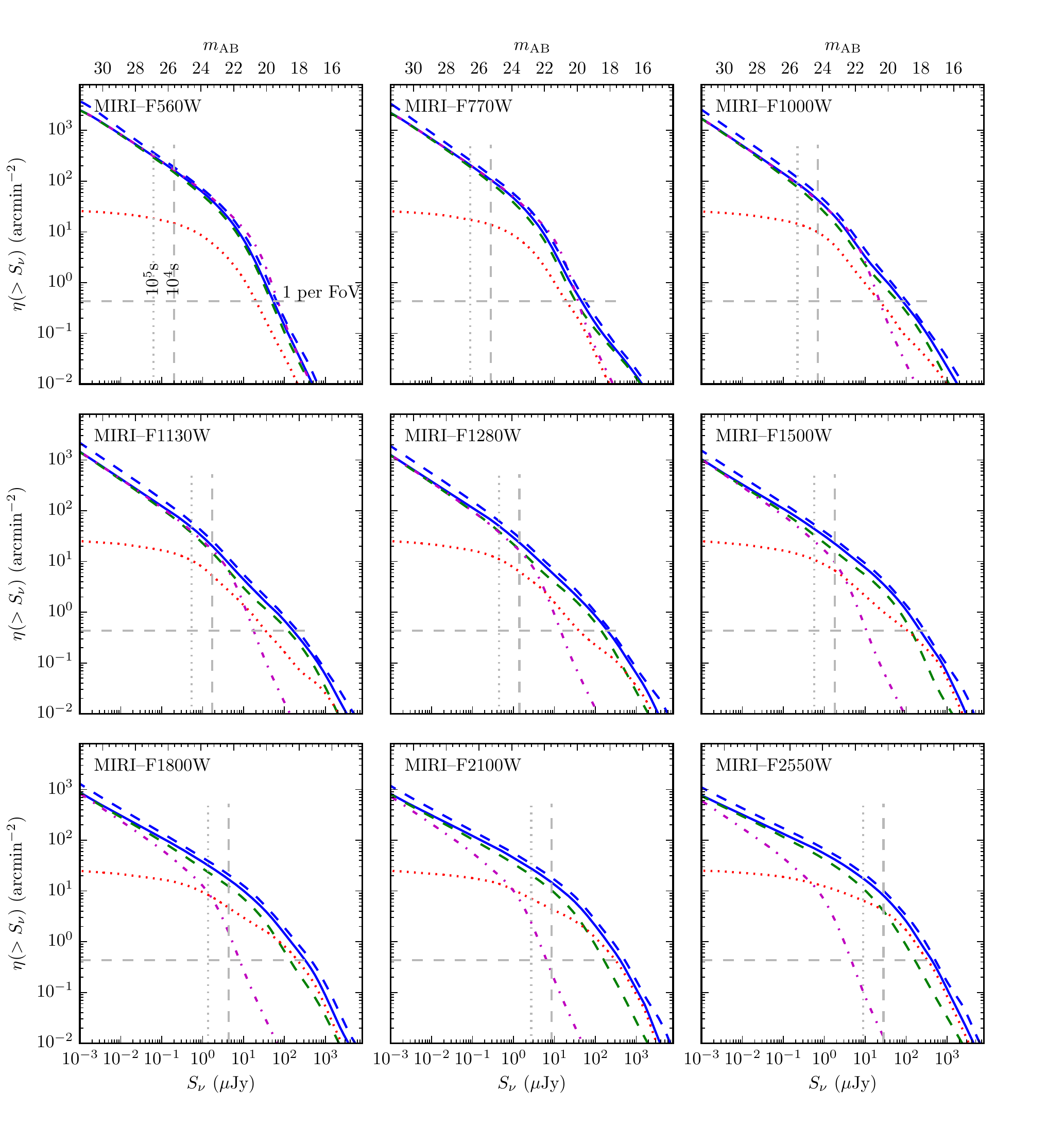}
\caption{Predicted cumulative galaxy number counts in the MIRI bands.  The name of the band is indicated each panel.  All lines have the same meaning as in Fig.~\ref{fig:cncts_NIRCam}.}
\label{fig:cncts_MIRI}
\end{figure*}
\begin{figure*}
\includegraphics[width = \linewidth]{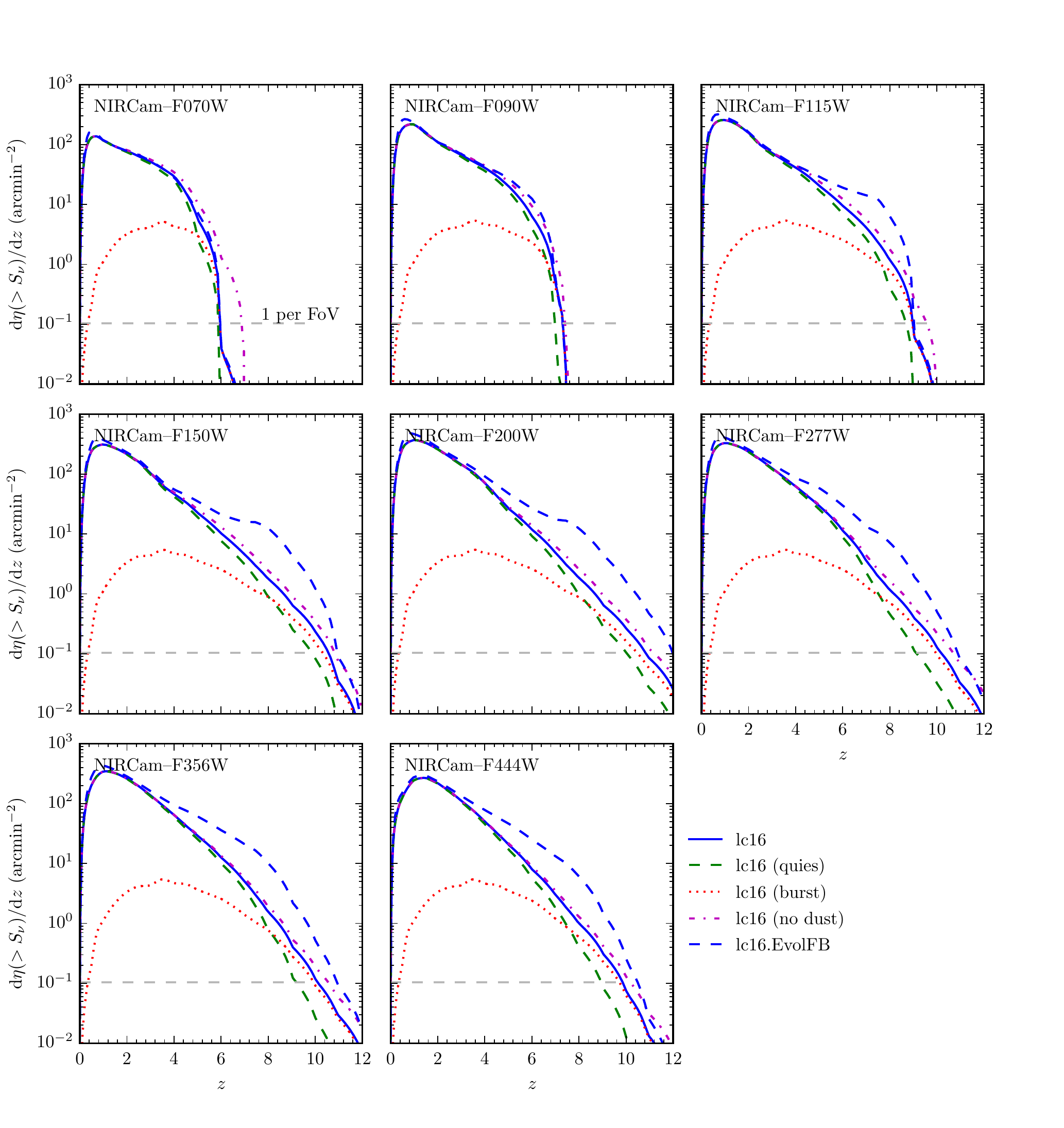}
\caption{Predicted redshift distributions for objects detectable in a $10^{4}$~s exposure in NIRCam bands.  The name of the band is indicated in each panel.  The solid blue lines show the predictions of the fiducial model, and the dashed green and dotted red lines show the contribution to this from quiescent and starburst galaxies respectively.  The predictions of the fiducial model excluding dust absorption are shown by the dash-dotted magenta lines.  The dashed blue lines show the predictions from the evolving feedback variant.  For reference, the horizontal dashed line indicates the number density at which there is one object per field of view per unit redshift.}
\label{fig:dndzs_NIRCam}
\end{figure*}
\begin{figure*}
\includegraphics[width = \linewidth]{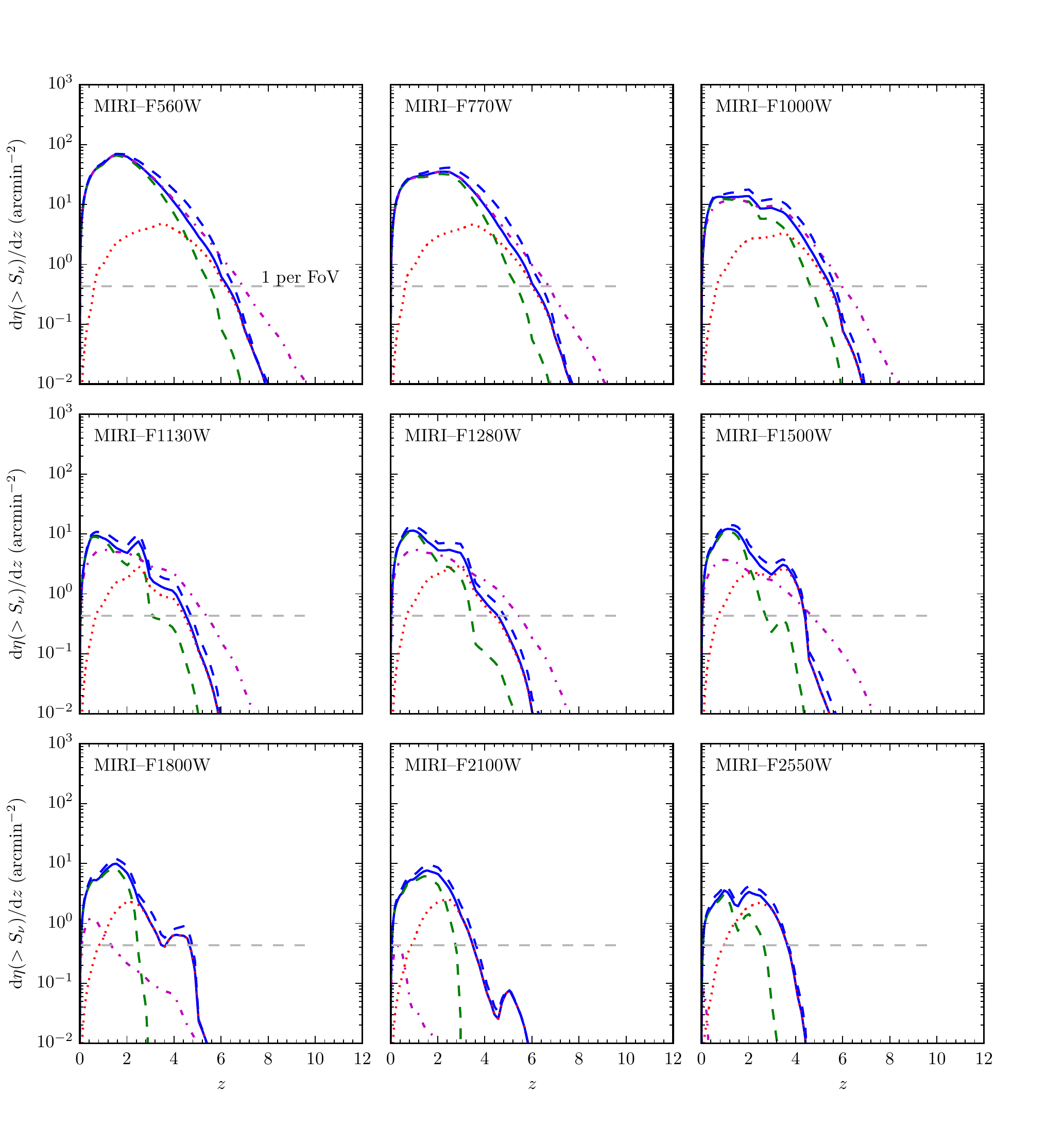}
\caption{Predicted redshift distributions for galaxies observable with a $10^{4}$~s exposure in MIRI bands.  The name of the band is indicated in each panel.  All lines have the same meaning as in Fig.~\ref{fig:dndzs_NIRCam}.}
\label{fig:dndzs_MIRI}
\end{figure*}
\begin{figure*}
\centering
\includegraphics[width=\linewidth]{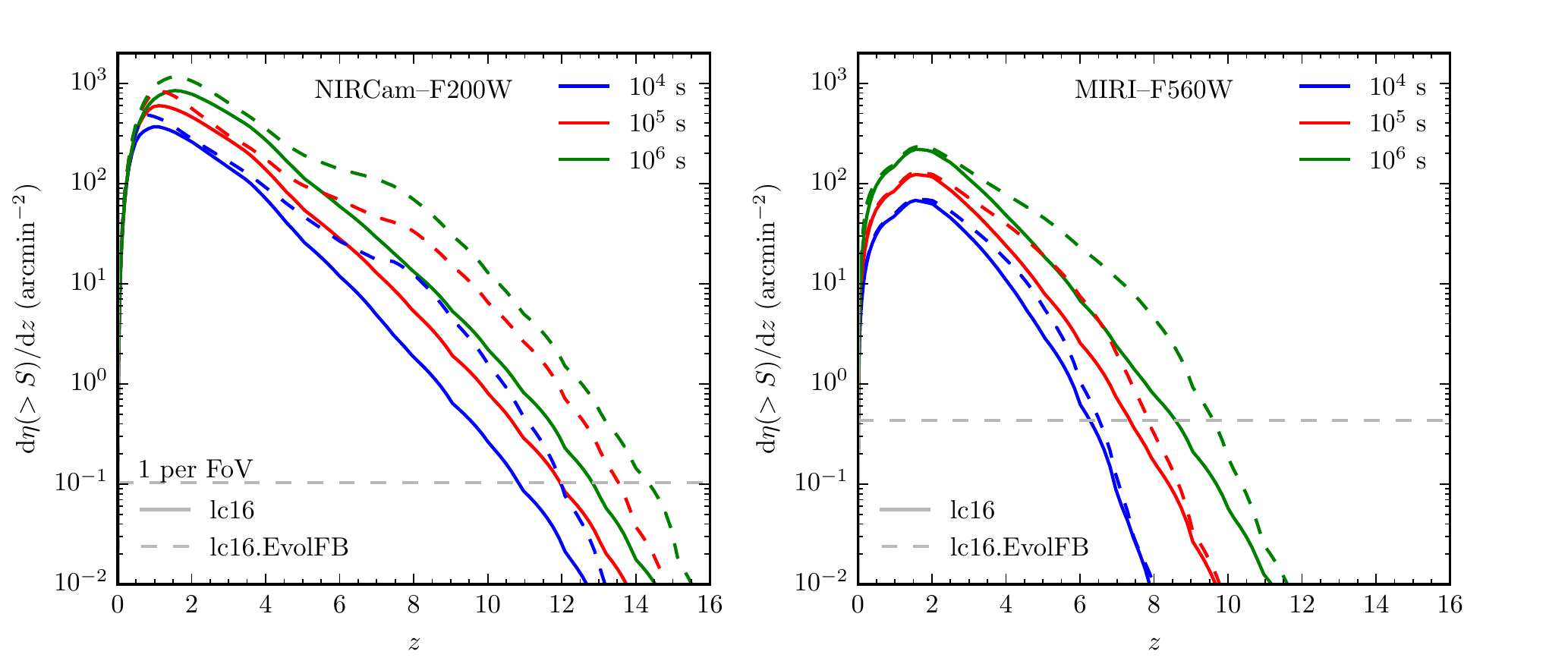}
\caption{Predicted redshift distributions for the NIRCam--F200W (left panel) and MIRI--F560W (right panel) bands for galaxies observable with a range of exposure times.  The blue, red and green lines show predictions for exposures of $10^{4}$, $10^{5}$ and $10^{6}$~s respectively.  The solid and dashed lines are the predictions of the fiducial and evolving feedback variant models respectively.  For reference, the horizontal dashed lines show the number surface density at which there is one object per field of view per unit redshift.}
\label{fig:dndz_limits}
\end{figure*}
The simplest statistic of a galaxy population that can be derived from an imaging survey is their number counts.  Here we present the predictions for the cumulative number counts observable with NIRCam (Fig.~\ref{fig:cncts_NIRCam}) and MIRI (Fig.~\ref{fig:cncts_MIRI}). We also show the corresponding redshift distributions (for a $10^4$~s exposure) in Fig.~\ref{fig:dndzs_NIRCam} (NIRCam) and Fig.~\ref{fig:dndzs_MIRI} (MIRI). We obtain the number counts and redshift distributions by integrating the predicted luminosity functions according to
\begin{equation}
\frac{\mathrm{d}^3 \eta}{\mathrm{d}\ln S_{\nu}\,\mathrm{d}z\,\mathrm{d}\Omega}=\frac{\mathrm{d}n}{\mathrm{d}\ln L_{\nu}}\frac{\mathrm{d}^{2}V}{\mathrm{d}z\,\mathrm{d}\Omega}\rm,
\end{equation}
where $\eta$ is the surface density of galaxies projected on the sky, $n$ is the number density of galaxies and $\mathrm{d}^{2}V/\mathrm{d}z\,\mathrm{d}\Omega$ is the comoving volume element per unit solid angle.
We show the contribution to the predicted number counts and redshift distributions from quiescent and starburst galaxies.  For the NIRCam filters, the counts are dominated by quiescent galaxies.  This is because they are dominated by galaxies at low redshift, for which starbursts are not a significant population at these wavelengths.  This is also why the predicted number counts from the fiducial and evolving feedback variant models are so similar, as at low redshifts the feedback normalisations are equal, though the lc16.EvolFB model does predict slightly more galaxies at faint fluxes.  For the MIRI number counts, we see the burst population becoming important at brighter fluxes in bands $\lambda_{\rm obs}\gtrsim10$~$\mu$m. These wavelengths also correspond to a shift from the number counts being dominated by dust-attenuated stellar light to dust emission.  Again, these number counts are dominated by relatively low-redshift galaxies, for which the MIRI filters probe the dust emission from the rest-frame mid-IR.

The redshift distributions in Figs~\ref{fig:dndzs_NIRCam} and~\ref{fig:dndzs_MIRI} exhibit a more discernible difference between the two models, particularly in the NIRCam bands at high redshift.  For instance, in the NIRCam--F200W filter, the redshift at which one object per FoV per unit redshift is predicted to be observable with a $10^4$~s exposure is $z\sim11$. For the evolving feedback variant $\sim5$ times more galaxies are predicted to be observable at this redshift.  From our predictions, it appears that very few galaxies will be observable at $z\gtrsim10$ with NIRCam and at $z\gtrsim6$ with MIRI, although we stress that this is the case for a single FoV and a $10^4$~s exposure.  Additionally, we note that we have not considered effects such as gravitational lensing, which would allow surveys to probe fainter galaxies at higher redshifts \citep[e.g.][]{Infante15}.  

Various features in the predicted MIRI redshift distributions can be related to PAH emission. For example, the peaks at $z\sim2.5$ in the MIRI--F1130W distribution and at $z\sim3.6$ in the MIRI-F1500W distribution correspond to the $3.3$~$\muup$m PAH feature.

We briefly consider the possibility that nebular emission lines may affect our predicted broadband photometry \citep[e.g.][]{Smit15}, as they are not included in our galaxy SEDs.  For this we focus on the MIRI--F560W filter at $z\sim7$ as the H~$\alpha$ emission line is redshifted across the filter.  The luminosity of the H~$\alpha$ line is calculated assuming that all photons emitted with wavelengths shorter than $912$~{\AA} will ionize a hydrogen atom in the gas surrounding the star.  We then assume `Case B' recombination i.e. we ignore recombinations directly to the ground state ($n=1$), as these just produce another ionizing photon.  Thus only recombinations to $n>1$ are counted.  The fraction of such recombinations that produce an H~$\alpha$ photon ($n=2\rightarrow1$) is taken from \cite{Osterbrock74}.  We apply the dust extinction factor predicted by \grasil at the wavelength of the line to the line luminosity.  We find that the predicted equivalent widths (EWs) of the line are $\sim400$~{\AA}, significantly narrower than the width of the MIRI--F560W filter $\sim1.2$~$\muup$m.  As a result, the line luminosity has a minor effect on the broadband photometry.  For example, at $z=7.5$ in both models $95$~per~cent of the sampled galaxies have their MIRI--F560W luminosity increased by less than $\sim10$~per~cent and $90$~per~cent by less than $\sim7$~per~cent.  This results in a negligible difference in the luminosity functions if H~$\alpha$ emission is included.  Thus we conclude that a more detailed inclusion of nebular emission lines \citep[e.g.][]{Panuzzo03} is unlikely to affect the results presented here (see also Bisigello et al. \citeyear{Bisigello16} for an investigation of the effect of nebular emission lines on MIRI photometry).

We now consider the predicted redshift distributions of galaxies that would be observable with longer exposures than considered in Figs~\ref{fig:dndzs_NIRCam} and \ref{fig:dndzs_MIRI}.  In Fig.~\ref{fig:dndz_limits} we show predictions for $10^{4}$, $10^{5}$ and $10^{6}$~s exposures, for the {NIRCam--F200W} and MIRI--F560W filters.  For the fiducial model a $10^{6}$~s exposure will increase the number of observable objects in the NIRCam--F200W filter at $z\sim11$ from $1$~per FoV to $\sim10$~per Fov, and will increase the highest redshift at which an object is observable in a single FoV from $z\sim11$ to $z\sim13$.  For the evolving feedback model, the highest redshift will be $z\sim14.5$.  Thus, we expect that long ($>10^4$~s) exposures with \emph{JWST} will provide better constraints on the effectiveness of supernova feedback in galaxies at high redshift.  
\subsection{Sizes of galaxies in \emph{JWST} bands}
\label{subsec:JWST_sizes}
\begin{figure*}
\centering
\includegraphics[width=\linewidth]{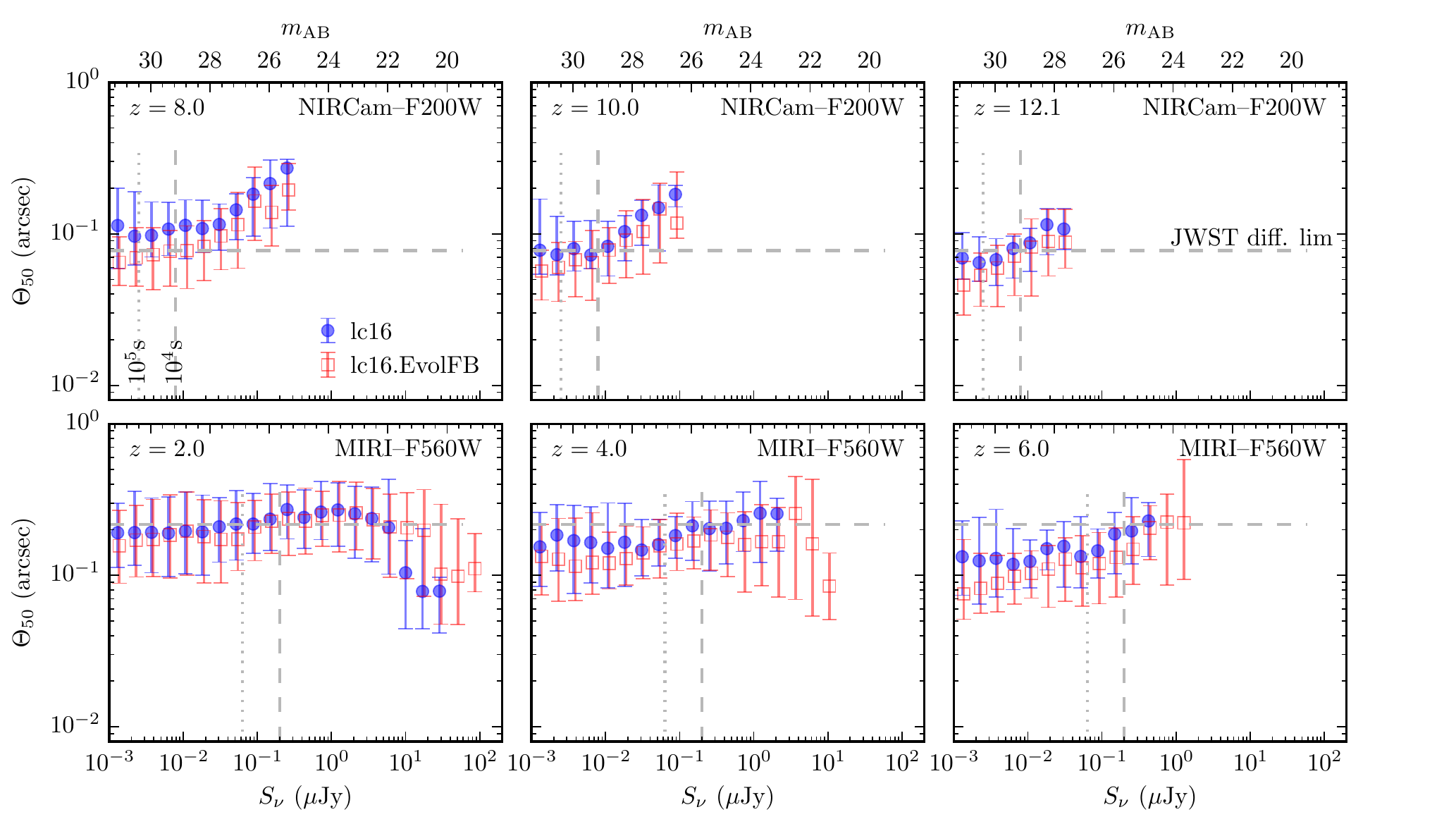}
\caption{Predicted projected half-light radii at a range of redshifts, as a function of galaxy flux, $S_{\nu}$.  The redshift is indicated in each panel.  The top row shows predictions for the NIRCam--F200W filter, the bottom row shows predictions for the MIRI--F560W filter.  The blue filled circles and open red squares respectively indicate the median size for the fiducial and evolving feedback variant models at a given flux, with the error bars indicating the 16-84 percentile ranges in a given flux bin.  For reference, the vertical dashed and dotted lines respectively indicate the sensitivity limit for a $10^{4}$ and $10^{5}$~s exposure for that filter.  The horizontal dashed line indicates the diffraction limit of \emph{JWST} for that filter, assuming a $6.5$~m diameter mirror.} 
\label{fig:sizes_JWST}
\end{figure*}
Finally, we present predictions for the angular sizes of galaxies for the NIRCam--F200W and MIRI-F560W filters in Fig.~\ref{fig:sizes_JWST}.  We make such predictions for all NIRCam and MIRI filters but show only these two here for brevity, the predictions for other filters will be made available online.  The sizes in each band are calculated as described in Section~\ref{subsec:L16_at_hi_z}.  

We can see that the predicted sizes are $\sim0.1$~arcsec, with the evolving feedback variant generally predicting slightly smaller sizes.  By comparison to the diffraction limits for \emph{JWST}, shown here as dashed horizontal lines, it is evident that NIRCam will be able to resolve the majority of detected galaxies whereas this will not be the case for MIRI (for $z\gtrsim2$).
\section{Summary}
\label{sec:conclusion}
The \emph{James Webb Space Telescope} (\emph{JWST}) is scheduled for launch in October 2018 and is expected to significantly advance our understanding of the high-redshift ($z\gtrsim7$) Universe.

Here we present predictions for deep galaxy surveys with \emph{JWST}.  To do so we couple the hierarchical galaxy formation model \galform\ \citep{Lacey16}, with the spectrophotometric code \grasil \citep{Silva98} for computing galaxy SEDs.  \grasil calculates the absorption and re-emission of stellar radiation by interstellar dust by solving the equations of radiative transfer in an assumed geometry.  This allows us to produce UV-to-mm galaxy SEDs, broadening the predictive power of the model to cover the full wavelength range that will be probed by \emph{JWST}. The galaxy formation model is implemented within a dark matter only $N$-body simulation using \emph{Planck} cosmological parameters \citep{PlanckCollab15}. Adjustable parameters in the model are calibrated against a broad range of observational data such as optical and near-IR luminosity functions at $z=0$, the evolution of the rest-frame near-IR luminosity functions for $z=0-3$, far-IR galaxy number counts and redshift distributions, and the evolution of the rest-frame far-UV luminosity function for $z=3-6$ (Lacey et al. \citeyear{Lacey16}; Baugh et al. in preparation).  Here we have shown that the model predicts evolution of the rest-frame far-UV luminosity function for $7\lesssim z\lesssim10$, and galaxy sizes for $7\lesssim z\lesssim9$, in good agreement with observations.  

We also present predictions for an evolving feedback variant model, in which the strength of supernova feedback is allowed to vary as a function of redshift \citep{HouJun16}.  This adjustment allows the model to reproduce the reionization redshift inferred from \emph{Planck} data \citep{PlanckCollab15}, as well as the $z=0$ luminosity function of the Milky Way satellites and their metallicity--stellar mass relation.   

We present predictions for \emph{JWST} in the form of luminosity functions, number counts, redshift distributions and angular sizes for each of the broadband filters on NIRCam and MIRI on \emph{JWST}, for both the fiducial model `lc16' and the evolving feedback variant `lc16.EvolFB'.

We find that for a $10^{4}$~s exposure the fiducial model predicts that \emph{JWST} will be able to observe a single galaxy per field of view at $z\sim11$ in the NIRCam--F200W filter; though the evolving feedback model predicts number surface densities factors of $\sim5$ greater.  The model predicts that similar exposures with MIRI will not detect any galaxies at $z\gtrsim6$ (in a single FoV).  Longer integration times will increase the number of galaxies that are observable, for example, a $10^6$~s integration will increase the number of galaxies predicted by the fiducial model to be observable in a single FoV by a factor of $\sim10$.  A similar effect may be achieved by utilising strong gravitational lenses; however, we do not consider such an effect here.  We consider a simple model for calculating H~$\alpha$ emission and conclude that nebular emission lines will have a negligible effect on these results.

The predicted sizes of high-redshift galaxies observable with \emph{JWST} are $\sim0.1$~arcsec, and as such we expect NIRCam to be capable of resolving the majority of detected galaxies.

We hope that the predictions presented here will help inform galaxy survey strategies for \emph{JWST}.  In the future, we plan to make our results public for such a purpose and to further develop our methodology to produce realistic mock galaxy catalogues for NIRCam and MIRI.  This will allow us to make direct predictions for the field-to-field variance.  We envisage that observations with \emph{JWST} will provide a wealth of information on physical processes important for galaxy formation, such as the effectiveness of supernova feedback in galaxies at high redshift.           
\section*{Acknowledgements}
We thank the anonymous referee for their comprehensive and constructive comments. The authors would like to thank Alessandro Bressan, Gian-Luigi Granato and Laura Silva for use of, and discussions relating to, the \grasil code. Additionally, we would like to acknowledge Karina Caputi, Lydia Heck, John Helly, Hou Jun, Sarah Kendrew, John Pye, Luiz Felippe Rodrigues, Martin Ward and Rogier Windhorst for many helpful discussions during the development of this work. This work was supported by the Science and Technology Facilities Council [ST/K501979/1, ST/L00075X/1]. CMB acknowledges the receipt of a Leverhulme Trust Research Fellowship. CSF acknowledges an ERC Advanced Investigator Grant, COSMIWAY [GA 267291] and the Science and Technology Facilities Council [ST/F001166/1, ST/I00162X/1].  This work used the DiRAC Data Centric system at Durham University, operated by the Institute for Computational Cosmology on behalf of the STFC DiRAC HPC Facility (www.dirac.ac.uk). This equipment was funded by BIS National E-infrastructure capital grant ST/K00042X/1, STFC capital grant ST/H008519/1, and STFC DiRAC Operations grant ST/K003267/1 and Durham University. DiRAC is part of the National E-Infrastructure. 
\bibliographystyle{mnras}
\bibliography{ref} % if your bibtex file is called example.bib
%\begin{thebibliography}{99}
%\end{thebibliography}
%%%%%%%%%%%%%%%%%%%%%%%%%%%%%%%%%%%%%%%%%%%%%%%%%%
%%%%%%%%%%%%%%%%% APPENDICES %%%%%%%%%%%%%%%%%%%%%
\appendix
%If you want to present additional material which would interrupt the flow of the main paper, it can be placed in an Appendix which appears after the list of references.
\section{Dependence of high-redshift predictions on model assumptions}
\label{sec:discussion}
\begin{figure*}
\centering
\includegraphics[width=\linewidth]{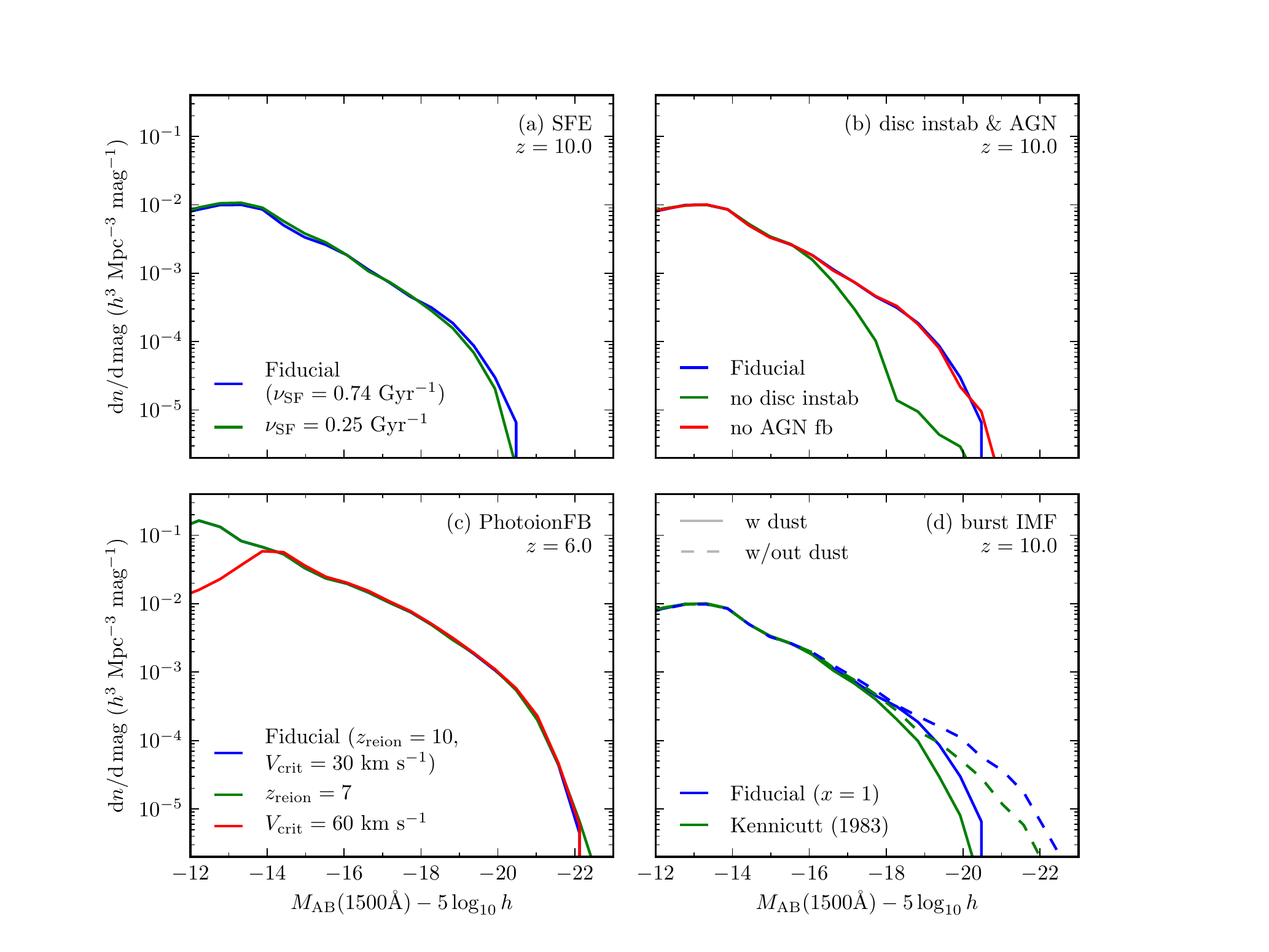}
\caption{Predictions from variant models for the rest-frame far-UV (1500~\AA) luminosity functions for $z=10$ [panels (a), (b) and (d)] and for $z=6$ [panel (c)]. \emph{Panel (a)}: the effect of varying the star formation efficiency parameter $\nu_{\rm SF}$.  \emph{Panel (b)}: the effect of turning off disc instabilities and AGN feedback. \emph{Panel (c)}: the effect of varying the photoionization feedback parameters: $z_{\rm reion}$ and $V_{\rm crit}$. \emph{Panel (d)}: the effect of varying the burst IMF both with (solid lines) and without dust attenuation (dashed lines).}
\label{fig:param_explore}
\end{figure*}
The galaxy formation model used in this work assumes prescriptions for the physical processes involved in galaxy formation that are in some cases calibrated on observational, or simulation, data at $z\lesssim6$. In this Appendix, we briefly discuss the dependence of our results on some of the assumptions made and their applicability at the high redshifts explored in this paper. We note that this is not intended to be an exhaustive discussion (which we consider to be beyond the scope of this work), and emphasize that the primary focus of this study is to present the predictions for the outcomes of future \emph{JWST} galaxy surveys for the model described in \cite{Lacey16}, rather than to investigate to what extent these predictions are sensitive to the various prescriptions for physical processes within the model. We envisage that this will form the basis of future study once the predictions presented earlier are confronted with the corresponding observations. 

Some results from a brief parameter space exploration at high redshift are shown in Fig.~\ref{fig:param_explore}, for a more detailed exploration of the parameter space of this model we refer to reader to Appendix~C of \cite{Lacey16}.  For ease of computation, the results presented in Fig.~\ref{fig:param_explore} have used the model for the absorption and re-emission of radiation by interstellar dust described in Lacey et al. (\citeyear{Lacey16}, see also Cowley et al., \citeyear{Cowley16SEDs}), rather than \grasil. However, for the rest-frame wavelength investigated in Fig.~\ref{fig:param_explore} ($1500$~\AA) the difference between the predictions of these two models is negligible.
\subsection{Formation of molecular hydrogen and star formation efficiency}
The model assumes a star formation law for galactic discs in which the surface density of star formation is related to the surface density of molecular gas (see equation~\ref{eq:star_formation_law}). The molecular gas fraction, $f_{\rm mol}$, is computed according to the mid-plane gas pressure in the disc using the empirical relation of \cite{BlitzRosolowsky06}. Any possible dependence of the molecular gas fraction on physical properties such as the gas metallicity or the interstellar far-UV radiation field as proposed in some models \citep[e.g.][]{Gnedin:2009,KMT:2009} are not explicitly considered. The model for the formation of molecular hydrogen of \cite{KMT:2009} was implemented into \galform\ by \cite{Lagos11}, however, Lagos et al. concluded that the \cite{BlitzRosolowsky06} relation produced better agreement with observations at low redshift.

The value for the inverse timescale of star formation from molecular gas, sometimes referred to as the star formation efficiency, $\nu_{\rm SF}$ (see equation~\ref{eq:star_formation_law}), is chosen based on the value measured for this parameter by \cite{Bigiel11}. Those authors presented the $1\sigma$ range for this quantity as being $0.25-0.74$~Gyr$^{-1}$, so in practice, this parameter is only varied within this range during the model calibration. Doing so has a negligible effect on our predictions for the rest-frame far-UV ($1500~{\AA}$) luminosity function at $z=10$ [see Fig.~\ref{fig:param_explore}(a)].
\subsection{Feedback processes}
Throughout this work, we have explored a variant model in which the strength of SN feedback is allowed to vary as a function of redshift. This variant allows the model to reproduce a higher redshift of reionization in better agreement with that inferred from \emph{Planck} data, as well as the luminosity function and metallicity--stellar mass relation of $z=0$ Milky Way satellite galaxies, though it does introduce some new parameters in the SN feedback model (Hou et al., \citeyear{HouJun16}).  We find that this modification increases the numbers of galaxies observable with \emph{JWST} at high-redshift by factors of $\sim5$.

The prescription for AGN feedback in our model was first introduced by \cite{Bower06} and has a minimal impact on model predictions at high redshifts. In this prescription, the conditions for gas cooling in the halo to be turned off as a result of thermal energy deposited by relativistic radio jets are: (i) that the gas is cooling quasi-statically (i.e. the cooling time is long compared to the free-fall time); and (ii) that the SMBH is massive enough such that the power required to balance the radiative cooling luminosity is below some fraction of its Eddington luminosity. These conditions are rarely met at high-redshift as: (i) gas cooling times are generally shorter as the physical density of the Universe is higher ($\tau_{\rm cool}\propto\rho^{-1}$); and (ii) typically the SMBHs have not had time to grow massive enough to satisfy the second criterion.  As such, turning off AGN feedback has a negligible impact on our prediction for the rest-frame UV luminosity function at $z=10$ [see Fig.~\ref{fig:param_explore}(b)]. However, we stress that turning off AGN feedback does not result in a viable model of galaxy formation, as it fails to reproduce many properties of the galaxy population at $z=0$ (see e.g. Fig.~C2 of Lacey et al., \citeyear{Lacey16}).

Photoionization feedback is implemented such that for $z<z_{\rm reion}$, no cooling of gas occurs in halos with circular velocities $V_{\rm c}<V_{\rm crit}$ \citep{Benson:2002,Font:2011}. Adjusting $z_{\rm reion}$ has a negligible impact on our high-redshift predictions at luminosities observable by JWST, and adjusting $V_{\rm crit}$ mainly shifts the `break' at the faint end of the luminosity function [see Fig.~\ref{fig:param_explore}(c)]. It should be noted, however, that these parameters are considered `fixed' (see Table~1 of Lacey et al., \citeyear{Lacey16}) and are not varied in the model calibration.  The value for $z_{\rm reion}$ is chosen based on \emph{WMAP7} data \citep{Dunkley:2009} and the value for $V_{\rm crit}$ is based on cosmological gas simulations \citep{Hoeft:2006,Okamoto:2008}.              
\subsection{Disc instabilities and top-heavy IMF}
The bright ends of the predicted NIRCam and MIRI luminosity functions presented in this work are predicted to be dominated by galaxies undergoing a disc instability-triggered starburst for $z\gtrsim2$, so they will be somewhat sensitive to the treatment of this process. For example, turning off disc instabilities shifts the bright end of the rest-frame far-UV luminosity function towards fainter magnitudes by $\sim2$ magnitudes [see Fig.~\ref{fig:param_explore}(b)]. We stress very strongly, however, that turning this process off does not result in a viable model of galaxy formation, as it no longer reproduces the observed far-IR/sub-mm galaxy number counts and redshift distributions, which are predominantly composed of disc-instability triggered starburst galaxies in the redshift range $z\sim1-4$ \citep{Lacey16}.

A related issue is the assumption of a mildly top-heavy IMF for star formation during a starburst episode. This assumption is made so that the model can reproduce the observed far-IR/sub-mm galaxy number counts and redshift distributions \citep{Baugh05, Cowley15, Lacey16}, though the slope for the IMF adopted here is much less top-heavy than was advocated by Baugh et al. Baugh et al. suggested a slope of $x=0$ in $\mathrm{d}N/\mathrm{d}\ln m\propto m^{-x}$, however in the updated \galform\ model used here, $x=1$ was found to produce better agreement with the observed far-IR/sub-mm galaxy number counts and redshift distributions \citep{Lacey16}. 

This assumption has only a fairly minor impact on our model predictions in the rest-frame far-UV, as is shown in Fig.~\ref{fig:param_explore}(d), where we compare the predictions of the fiducial model to a model that adopts a \cite{Kennicutt83} IMF for both disc and burst star formation (the fiducial model adopts a Kennicutt IMF for disc star formation only). The increase in intrinsic UV flux as a result of forming more massive stars is counterbalanced by the increase in dust mass (which absorbs the UV radiation) due to the increased rate of SNe returning metals to the ISM. We emphasize this point by also showing the model predictions without dust attenuation in this panel, the difference between the two predicted luminosity functions is greater without dust extinction. As a result, the rest-frame far-UV luminosity function is fairly insensitive to the assumption of a top-heavy IMF for burst star formation, which has a much greater effect in far-IR/sub-mm bands that trace the dust emission, against which the model has been calibrated.
\subsection{Stellar population synthesis model}
The models presented here make use of the stellar population synthesis (SPS) model of \cite{Maraston05}. This includes an empirical calibration for the light produced by thermally-pulsating asymptotic giant branch stars (TP-AGB). This is difficult to model accurately from purely theoretical stellar evolution models, so Maraston (\citeyear{Maraston05}) calibrate this using observations of star clusters. This phase is important for the rest-frame near-IR luminosities of stellar populations with ages $\sim0.1-1$~Gyr. We refer the interested reader to \cite{ConroyARAA13} for a review of the Maraston, and other, SPS models. A comprehensive investigation of the effects of implementing different SPS models within \galform\ was performed by \cite{vgp14}, who found that SPS models with a larger contribution from TP-AGB stars produced stronger evolution in the rest-frame near-IR luminosity function. \cite{Lacey16} found that implementing different SPS models made only a modest change to the predicted rest-frame $K$-band luminosity function, but that the Maraston (\citeyear{Maraston05}) models produced better agreement with observations of this quantity up to $z\sim3$ (see Fig.~C16 of Lacey et al., \citeyear{Lacey16}).

Recently, \cite{Stanway:2016} proposed an SPS model that includes the effects of binary stars ({\sc bpass}), which produces higher hydrogen-ionizing luminosities than single-star SPS models by up to $\sim60$~per~cent at sub-solar ($0.1-0.2$~Z$_{\odot}$) metallicities.  \cite{Wilkins:2016} found that the increased ionizing radiation that results from assuming this SPS model also increases the nebular emission by up to $\sim25$~per~cent. It is conceivable that assuming the Stanway et al. models would have a similar effect on predictions from our simple model for nebular emission. However, given that we estimate the impact of nebular emission on our predictions for the broadband luminosity functions presented in this work to be negligible, it is unlikely that an increase of only $25$~per~cent in nebular emission would produce a significant effect on our predicted broadband luminosity functions.
%%%%%%%%%%%%%%%%%%%%%%%%%%%%%%%%%%%%%%%%%%%%%%%%%%
% Don't change these lines

\bsp	% typesetting comment
\label{lastpage}
\end{document}